\documentclass[fleqn,usenatbib]{mnras}

\usepackage{newtxtext,newtxmath}

\usepackage[T1]{fontenc}

\DeclareRobustCommand{\VAN}[3]{#2}
\let\VANthebibliography\thebibliography
\def\thebibliography{\DeclareRobustCommand{\VAN}[3]{##3}\VANthebibliography}

\usepackage{graphicx}	
\usepackage{amsmath}	

\usepackage{multirow}

\title[Tracing mass-loss history of NGC 3132]{Dissecting NGC 3132: Tracing the mass-loss history of the southern ring planetary nebula}

\author[Bouvis et al.]{
Bouvis K.$^{1,2}$\thanks{E-mail: kbouvis@noa.gr},
Akras S.$^{1}$,
Monteiro H.$^{3,4}$,
Konstantinou L.$^{1,2}$,
Boumis P.$^{1}$,
Garc\'{\i}a-Rojas J.$^{5,6}$,
\newauthor{Gon\c{c}alves D. R.$^{7}$,
Monreal-Ibero A.$^{8}$,
Aleman I.,$^{9}$,
and Gourgouliatos K. N.$^{2}$}
\\
$^{1}$Institute for Astronomy, Astrophysics, Space Applications and Remote Sensing, National Observatory of Athens, GR 15236 Penteli, Greece\\
$^{2}$Department of Physics, University of Patras, Patras, 26504 Rio, Greece\\
$^{3}$School of Physics and Astronomy, Cardiff University, Queen's Buildings, The Parade, Cardiff CF24 3AA, UK\\
$^{4}$Instituto de F\'{i}sica e Qu\'{i}mica, Universidade Federal de Itajuba, Av. BPS 1303-Pinheirinho, 37500-903, \'{I}tajuba, Brazil\\
$^{5}$Instituto de Astrof\'isica de Canarias, E-38205 La Laguna, Tenerife, Spain\\
$^{6}$Departamento de Astrof\'isica, Universidad de La Laguna, E-38206 La Laguna, Tenerife, Spain\\
$^{7}$Observat\'orio do Valongo, Universidade Federal do Rio de Janeiro, Ladeira Pedro Antonio 43, 20080-090, Rio de Janeiro, Brazil\\
$^{8}$Leiden Observatory, Leiden University, P.O. Box 9513, 2300 RA Leiden, The Netherlands\\
$^{9}$Laborat\'{o}rio Nacional de Astrof\'{i}sica, Rua dos Estados Unidos, 154, Bairro das Na\c{c}\~{o}es, Itajub\'{a}, MG, 37504-365, Brazil
}

\date{Accepted XXX. Received YYY; in original form ZZZ}

\pubyear{\the\year{}}

\begin{document}
\label{firstpage}
\pagerange{\pageref{firstpage}--\pageref{lastpage}}
\maketitle

\begin{abstract}
Central to our understanding of stellar evolution and its impact on processes in our Galaxy and across the universe is the study of mass loss. While the general framework is well established, recent \textit{JWST} observations of objects like NGC~3132 have revealed intricate nebular structures, suggesting complex mass-loss processes likely driven by multiple star system at its core. These findings pose new challenges for the currently available investigation tools. The primary goal of this study is the first detailed comparison of the physical properties and chemical composition obtained for NGC~3132, based on the latest detailed 3D model and observations from MUSE, \textit{JWST} and \textit{Spitzer}. We evaluate the reliability of the traditional empirical method and photoionization model for abundances estimations, both based on the same available high-quality, spatially resolved observations. We find that the model and empirical method yield consistent results for the integrated total properties such as $T_{\rm e}$, $n_{\rm e}$ and chemical abundances. However, when applied to simulated observations from the model, the empirical method fails to recover the model input abundances, providing only an approximate estimate. This discrepancy arises in part from the loss of information when summing fluxes over regions which have complex ionisation structures. This discrepancy in the case of oxygen has been estimated to be up to 35$\%$. Moreover, the latest IR data reveal a spatial correlation between H$_2$, c(H~$\upbeta$) as well as the [8.0]/[4.5] IRAC ratio. Finally, new clumps are discovered in [Ni~{\sc ii}] 7378~\AA, [Fe~{\sc ii}] 8617~\AA~and [Fe~{\sc iii}] 5270~\AA~emission lines.
\end{abstract}




\section{Introduction}

\begin{figure}
    \centering
    \includegraphics[width=0.75\columnwidth]{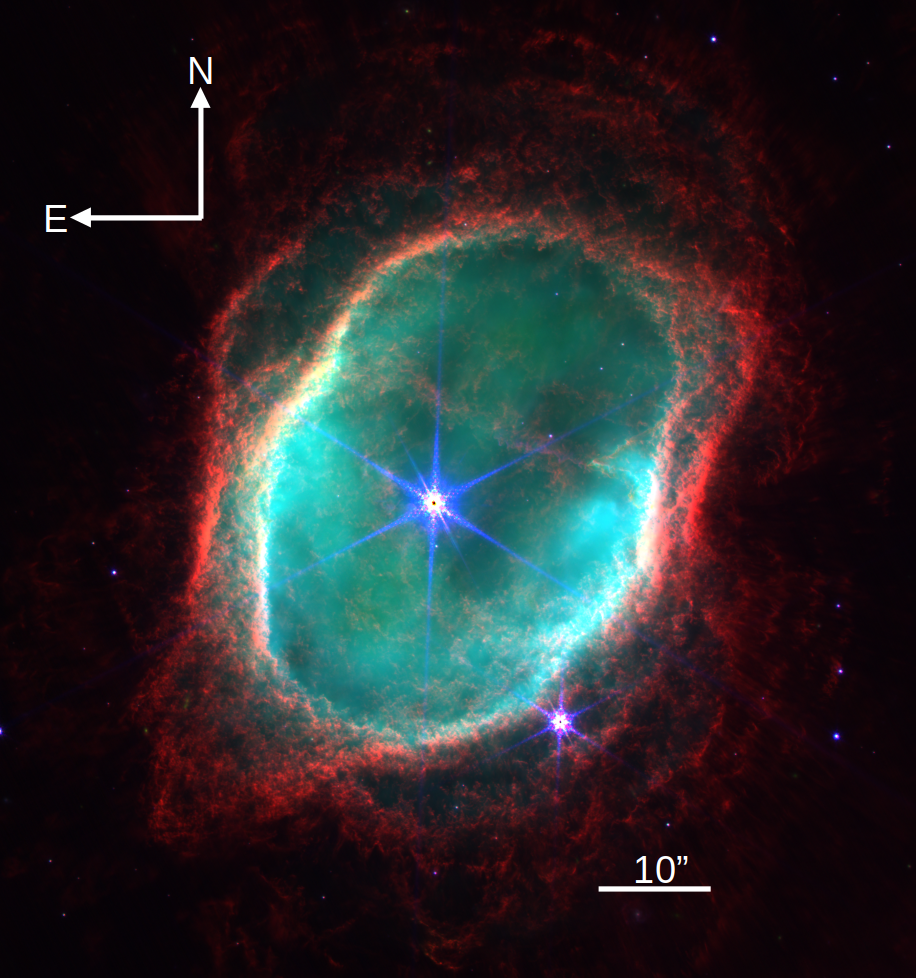}
    \caption{\textit{JWST} NIRCam composite image of NGC~3132. In the image, the RGB channels represent, respectively, the following emission dominantly: Red: H$_2$ 2.12~$\upmu$m, Green: Br~$\upalpha$ (4.05 $\upmu$m), Blue: [S~{\sc iii}] 9069~\AA.}
    \label{ngc3132_jwst_rgb}
\end{figure}

The young planetary nebula (PN) NGC~3132 (Fig.~\ref{ngc3132_jwst_rgb}) is known to host a multiple stellar system at its core. Specifically, the companion star HD 87892 is an evolved main sequence star of spectral type A2V, characterized by an effective temperature $T_{\rm eff}$=9\,200 K, luminosity L$\sim$57 L$_{\odot}$ and a mass of $\sim$2.4 M$_{\odot}$ \citep{Demarco2022}. In contrast, the pre-white dwarf (pre-WD), with spectral type A0, has an estimated mass of $\sim$0.7 M$_{\odot}$ \citep{sahai2023}. The effective temperature and luminosity of the pre-WD have been determined as $T_{\rm eff}$=100\,000~K and L$\sim$155 L$_{\odot}$, respectively \citep{frew2008}. The orbital period of the binary system is 25\,500 years \citep{ciardullo1999}, with an angle of 45$\degr$ between the orbital plane and the plane of the sky \citep{sahai2023}.

Dynamical interactions between the pre-WD, which we believe generate the PN, and its stellar companions have shaped the PN NGC~3132 into an elliptical-bipolar structure. Material exchange in the central region likely contributes to the highly asymmetric molecular content observed in the nebula \citep{decin2020}. The lower estimate on the neutral mass of the nebula is 0.02 M$_{\odot}$ \citep{sahai1990}, while the estimation for the total molecular mass range from 0.015 to 0.150 M$_{\odot}$ \citep{kastner2024}, which is comparable to or even exceeds its ionised mass \citep[$\sim$0.027 M$_{\odot}$,][]{sahu1986}. Recent studies using \textit{JWST} and ALMA data further confirm that NGC~3132 is a molecule-rich nebula, detecting species such as H$_2$, CO, and CN \citep{Demarco2022,kastner2024}. 

The distance of the companion A-type star has been computed D = 754$_{-18}^{+15}$ pc (with lower and upper 1$\upsigma$-like confidence intervals) by inverting its parallax ($1.3198 \pm 0.0344$ mas) from GAIA DR3 database \citep{gaia,Chornay2021,BaylerJones2021}. Given the significant uncertainties in the distance to the central star of the planetary nebula (CSPN), the distance to the central star’s visual companion is adopted as the distance to the PN itself \citep{Demarco2022}.

In this work, we present the first detailed spatial comparison between MUSE observations and predictions from a sophisticated 3D photoionisation model of NGC~3132. To investigate the nebula's molecular structure, which is not included in the 3D model, we conducted a multi-wavelength analysis of NGC~3132 by combining optical data from MUSE with infrared observations from \textit{JWST} and \textit{Spitzer}. Furthermore, we identified faint emission lines, rarely observed in planetary nebulae, in the MUSE datacube.

Details on the MUSE, \textit{JWST}, and \textit{Spitzer} observations of NGC~3132 are provided in Sect.~\ref{observations}. Sections~\ref{sec3} and \ref{sec4} present the main results of this study, including a detailed spectroscopic comparison between the observed MUSE data and the results from the 3D photoionisation model, along with an evaluation of the empirical method for estimating elemental abundances. The radial analysis between the MUSE, \textit{JWST}, and \textit{Spitzer} data is discussed in Sect~\ref{JWST_results}. The discovery of new Ni(Fe)-rich clumps in NGC~3132 is presented in Sect~\ref{nife}. Finally, our conclusions are summarized in Sect.~\ref{summary}.

\section{Observational and modelled data}
\label{observations}
\subsection{MUSE data}

Our analysis is based on observational data from the Multi Unit Spectroscopic Explorer \citep[MUSE,][]{bacon2010}, an IFU instrument which is mounted on the Very Large Telescope (VLT), in Chile. MUSE observations of NGC~3132 were conducted on February 19, 2014 \citep{bacon2014,Ibero2020}, and the data were released from the MUSE commissioning observations at the VLT Yepun (UT4) telescope under Program ID 60.A-9100(A). MUSE operated in the nominal Wide Field Mode (WFM-NOAO-N), covering a sky area of 1\arcmin $\times$ 1\arcmin with 0.2\arcsec pixel scale and 1.25~\AA~spectral sampling, in a wavelength range of 4750-9300~\AA. At the time of observation, FWHM of the central star ranged from 0.7\arcsec at 9100~\AA~to 0.8\arcsec at 4800~\AA. A total of three pointings were combined in order to cover the full extent of the nebula ($\sim$1\arcmin $\times$ 2\arcmin), and the combined datacube has a total exposure time of t$_{\rm exp}$ = 600~s in the central region and t$_{\rm exp}$ = 180~s at the edges. Lastly, the emission line maps are extracted from the datacube by fitting a Gaussian profile and subtracting the continuum emission \citep[][and references therein]{jorge2022}.

\subsection{\textit{JWST} data}

In addition to MUSE data, recent observations of NGC~3132 from the James Webb Space Telescope (\textit{JWST}), obtained in 2022, were utilized (P.I. Pontoppidan K., program number: 2733). Specifically, \textit{JWST} Near Infrared Camera (NIRCam) observations \citep{Pontoppidan2022,Demarco2022} were conducted on March 6, 2022, followed by Mid-Infrared Instrument (MIRI) observations on June 12, 2022. MIRI \citep{miri} and NIRCam \citep{nircam} are the two \textit{JWST} imaging instruments, with wavelength coverages of 0.6-5 $\upmu$m and 5-28 $\upmu$m, respectively. NIRCam has a field of view of 2.2\arcmin $\times$ 2.2\arcmin, while MIRI covers 74\arcsec $\times$ 113\arcsec. Each instrument has a different angular resolution: NIRCam's Short Wavelength Channel (F090W, F187N and F212N) has a pixel scale of 0.031\arcsec, NIRCam's Long Wavelength Channel (F356W, F405N and F470N) has a pixel scale of 0.063\arcsec, while MIRI has a pixel scale of 0.11\arcsec. Tables~\ref{jwst_nircam_filters} and \ref{jwst_miri_filters} provide the characteristics, FWHM at the time of observation and the most prominent emission features covered by the NIRCam and MIRI filters used for observing NGC~3132, respectively.

\begin{table}
\centering
\caption{NIRCam filters used for NGC~3132 observations.}
\resizebox{0.5\textwidth}{!}{%
\begin{tabular}{|c|c|c|c|c|c|}
\hline
Instrument              & Filter  & $\uplambda_{0}$ & $\Delta \uplambda$ & FWHM & Usage \\ 
                   &         & ($\upmu$m)     & ($\upmu$m)     & (\arcsec) & \\ \hline
\multirow{6}{*}{NIRCam} & F090W   &  0.901 & 0.194 & 0.062& [S~{\sc iii}] \\  
                        & F187N   &  1.874 & 0.024 & 0.062& Pa~$\upalpha$ \\ 
                        & F212N   &  2.12  & 0.027 & 0.062& H$_{2}$       \\ 
                        & F356W   &  3.563 & 0.787 & 0.126& H$_{2}$, PAHs  \\ 
                        & F405N   &  4.055 & 0.046 & 0.126& Br~$\upalpha$ \\  
                        & F470N   &  4.707 & 0.051 & 0.126& H$_{2}$       \\ \hline
\end{tabular}
}
\label{jwst_nircam_filters}
\end{table}

\begin{table}
\centering
\caption{MIRI filters used for NGC~3132 observations. PAHs is an abbreviation for Polycyclic Aromatic Hydrocarbons.}
\resizebox{0.5\textwidth}{!}{%
\begin{tabular}{|c|c|c|c|c|c|}
\hline
Instrument              & Filter  & $\uplambda_{0}$ & $\Delta \uplambda$ & FWHM & Usage \\ 
                   &         & ($\upmu$m)     & ($\upmu$m)     & (\arcsec) & \\ \hline
\multirow{4}{*}{MIRI}   & F770W   &  7.7   & 2.2   & 0.29& H$_{2}$, PAHs  \\  
                        & F1130W  &  11.3  & 0.7   & 0.40& PAHs          \\  
                        & F1280W  &  12.8  & 2.4   & 0.44& [Ne~{\sc ii}] \\  
                        & F1800W  &  18.0  & 3.0   & 0.58& [S~{\sc iii}] \\ \hline
\end{tabular}
}
\label{jwst_miri_filters}
\end{table}

\subsection{\textit{Spitzer} data}

The \textit{Spitzer} Space Telescope was an infrared space telescope equipped with the Infrared Array Camera \citep[IRAC,][]{irac}. IRAC operated simultaneously in four different bands: 3.6 $\upmu$m, 4.5~$\upmu$m, 5.8 $\upmu$m, and 8.0 $\upmu$m, covering a field of view of 5.2\arcmin $\times$ 5.2\arcmin~with 1.2\arcsec~pixel scale. However, during the data reduction process, the images were resampled to a pixel size of 0.6\arcsec. NGC~3132 was observed by \textit{Spitzer} on 14 January 2009, under program ID 501179 (P.I.: Sellgren, Kris).

The third and fourth bands (Table~\ref{table:3}) have been considered  indicators of molecular gas, with stellar objects showing excess emission in these channels often referred to as "green fuzzies" \citep[][and references therein]{phillips2010,Akrasgreen}. 

\begin{table}
\centering
\caption{\textit{Spitzer} space telescope IRAC's channels characteristics.}
\resizebox{0.47\textwidth}{!}{%
\begin{tabular}{|c|c|c|c|c|} 
 \hline
 Channel & Effective $\uplambda$ & Bandwidth & FWHM & Usage \\ [0.1ex] 
 
  & ($\upmu$m) & ($\upmu$m) & (\arcsec) & \\
 \hline
 1 & 3.551 & 0.750 (21$\%$) & 1.44 & H$_2$ \\
 
 2 & 4.493 & 1.010 (23$\%$) & 1.43 & H {\sc i}, H$_2$\\

 3 & 5.730 & 1.420 (25$\%$) & 1.49 & H$_2$\\

 4 & 7.873 & 2.930 (37$\%$) & 1.79 & H$_2$, [Ar {\sc iii}]\\
 \hline
\end{tabular}
}
\label{table:3}
\end{table}

\subsection{3D photoionisation model by MOCASSIN}

For a comprehensive spectroscopic analysis of NGC~3132, we also perform a thorough comparison of the MUSE observations with the simulated observations from a 3D photoionisation model (emission lines maps) created with {\sc mocassin} \citep[MOnte CArlo SimulationS of Ionised Nebulae,][]{MOCASSIN} code. The simulated images have a spatial resolution of 0.89\arcsec~per pixel, and the full description of the model is presented in \citet{hektor2025}.

\section{Evaluating Simulated Observations}
\label{sec3}
As previously mentioned, a key objective of this study is to perform a detailed comparison of the physical properties and chemical composition derived for NGC~3132, using both the latest 3D photoionization model \citep{hektor2025} and high-quality MUSE observational data \citep{Ibero2020}. To this end, we leverage the ability to generate simulated observations from the model.

For the spectroscopic analysis of NGC~3132 model and MUSE data, the {\sc satellite} Python code \citep{akras_sat,akras2022may} was implemented. Four different modules are available in the {\sc satellite} code for conducting a comprehensive analysis of extended sources:
\begin{enumerate}
    \item Angular Analysis Module
    \item Radial Analysis Module
    \item Specific Slit Analysis Module
    \item 2D Analysis Module
\end{enumerate}
Several physical properties are computed by the code in 1D and 2D spaces, such as interstellar extinction c(H~$\upbeta$), electron temperature ($T_{\rm e}$), electron density ($n_{\rm e}$), ionic and total abundances. In addition, ionisation Correction Factors (ICFs)--necessary for the estimation of total elemental abundance--and emission lines ratios for determining the excitation mechanisms (e.g., photoionisation, shocks), are also calculated. In this study, ICFs and emission line diagnostics were applied with caution to long-slit data and were not used on spatially resolved data, as both are defined for integrated spectra of entire nebulae. The errors produced in the code are derived by a Monte Carlo approach. For the aforementioned estimations, {\sc PyNeb} 1.1.19 package \citep{luridiana2015, morisset2020} is utilized, with 'PYNEB\_21\_1' atomic data set.

\subsection{Electron temperature and density}

\begin{figure}
    \centering
    \includegraphics[width=0.49\textwidth]{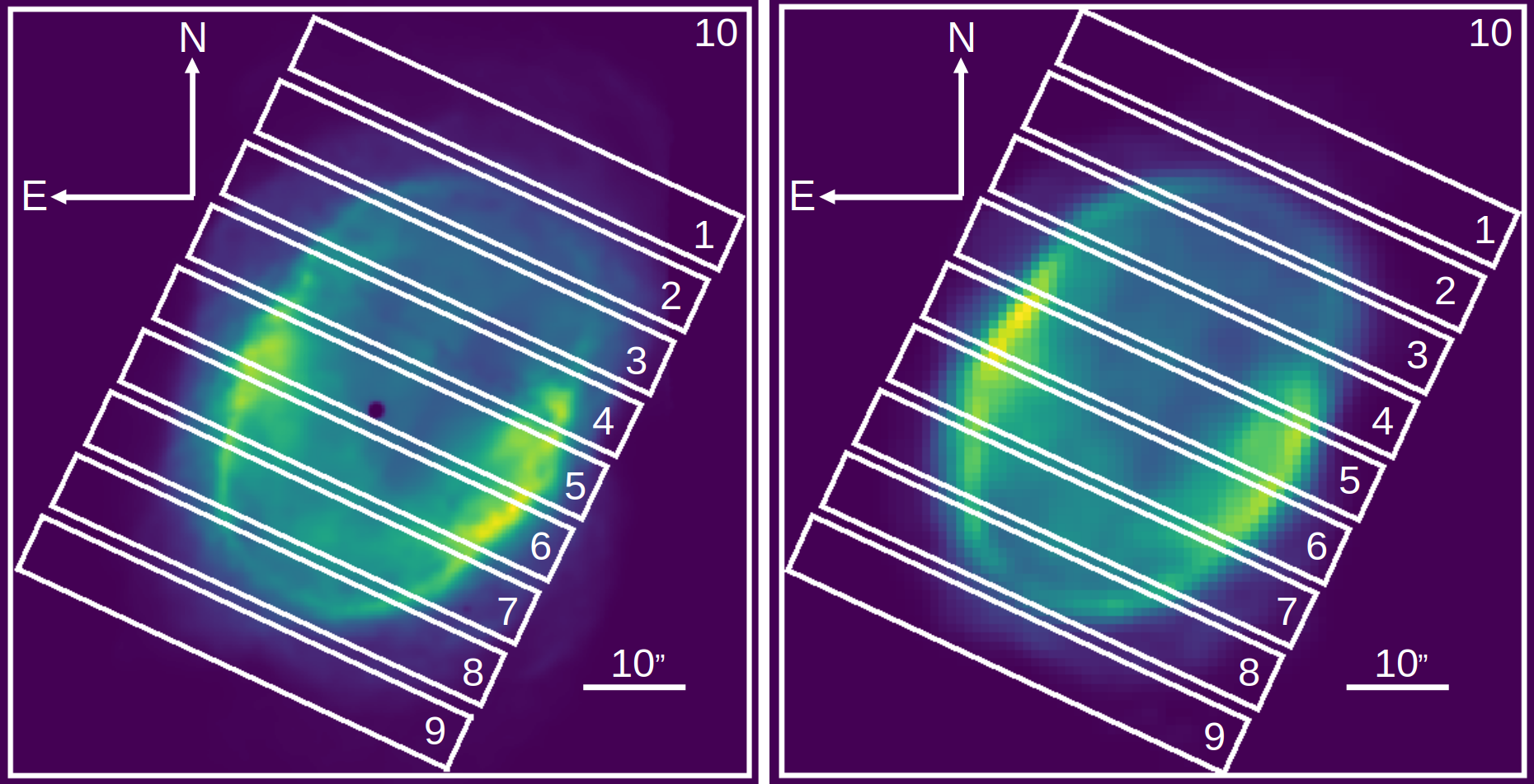}
    \caption{Ten pseudo-slits overlaid on the H~$\upalpha$ map of NGC~3132 as captured from the MUSE (left panel) and the photoionisation model (right panel).}
    \label{SAT_spec}
\end{figure}

The {\sc satellite} code allows us to perform the first detailed comparison of an extended, resolved nebula, such as NGC~3132, with the predictions of a 3D photoionisation model in a 2D spatial context. Nine pseudo-slits scanning the entire nebula from northwest to southeast (no. 1 to 9) and one covering the whole nebula (no. 10) were selected (Fig.~\ref{SAT_spec}). This approach aims to analyse specific regions of the nebula and the dependence of its physicochemical properties on the slit position, as well as to assess the consistency of the model.

\begin{figure}
    \centering
    \includegraphics[width=0.45\textwidth]{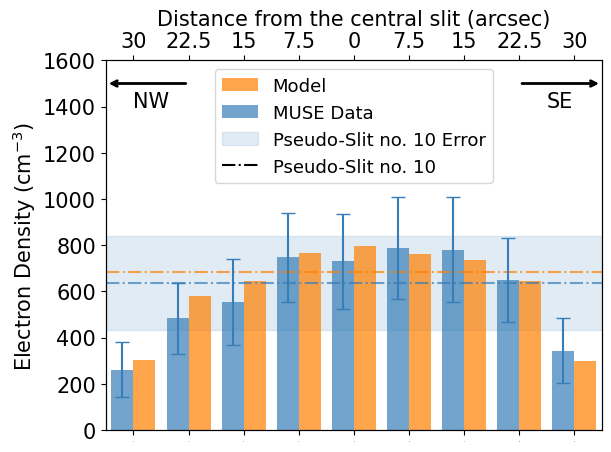}
    \includegraphics[width=0.45\textwidth]{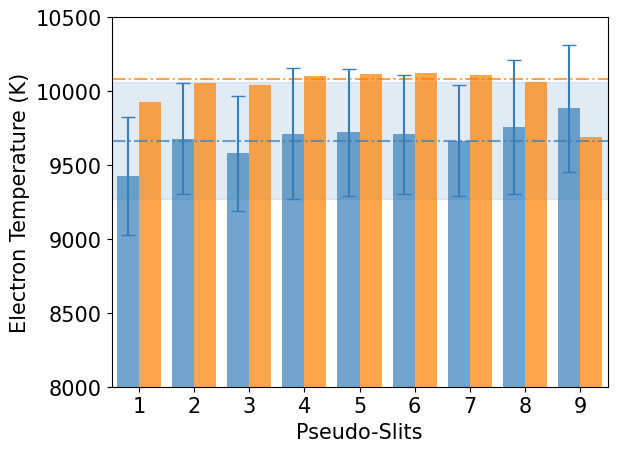}
    \caption{$n_{\rm e}$ [S~{\sc ii}] (top panel) and $T_{\rm e}$ [N~{\sc ii}] (bottom panel) for each pseudo-slit. Blue colour denotes the MUSE data, while orange the model values. The dashed lines show the values for the entire nebula (slit no. 10) and the light blue regions indicate the error of the observed measurement.} 
    \label{Ne_Te_spec}
\end{figure}

Physicochemical parameters, such as electron temperature, electron density, and chemical abundances, were computed for all ten pseudo-slits with the same methodology for both observed and simulated data. Specifically, for the low-ionisation plasma, including He$^+$, O$^+$, N$^+$ and S$^+$ ions, $n_{\rm e}$ was estimated from the [S~{\sc ii}] $\lambda\lambda$6716, 6731 ratio, and $T_{\rm e}$ from the [N~{\sc ii}] $\lambda\lambda\lambda$5755, 6548, 6584 ratio, while for moderate to high ionisation plasma, including He$^{+2}$, O$^{+2}$, S$^{+2}$, Ar$^{+2}$ and Cl$^{+2}$ ions, $n_{\rm e}$ was estimated from [Cl~{\sc iii}]~$\lambda\lambda$5517, 5538 ratio and $T_{\rm e}$ from [S~{\sc iii}] $\lambda\lambda$6312, 9069 ratio. Figure~\ref{Ne_Te_spec} shows the distribution of $n_{\rm e}$ [S~{\sc ii}] (upper panel) and $T_{\rm e}$ [N~{\sc ii}] (lower panel) derived from the observed (blue bars) and modelled (orange bars) emission lines maps. The model predicts $n_{\rm e}$ that closely match those derived from the MUSE data. Both the observations and the model show lower values at the outer regions (pseudo-slits no. 1, 2, 8, 9) and higher values at the inner region (pseudo-slits no. 4, 5, 6, 7). Regarding $T_{\rm e}$ [N~{\sc ii}], the model slightly overestimates the values, thought they remain within the uncertainties for most of the pseudo-slits. Table~\ref{table_te_ne} lists $n_{\rm e}$ and $T_{\rm e}$ derived from different plasma diagnostics based on the integrated observed and modelled spectra of the entire nebula (pseudo-slit no. 10). Overall, these values show good agreement mostly within the uncertainties. The corresponding plots for $T_{\rm e}$ [S~{\sc iii}] and $n_{\rm e}$ [Cl~{\sc iii}] are provided in Fig.~\ref{ne_cl3}.

\begin{table}
\centering
\caption{$T_{\rm e}$ and $n_{\rm e}$ estimated for pseudo-slit no. 10, covering the whole nebulae both in MUSE and modelled data employing the {\sc satellite} code.}
\resizebox{0.34\textwidth}{!}{%
\begin{tabular}{|c|c|c|}
\hline
Parameter & Model & MUSE \\ \hline
$T_{\rm e}$ [N~{\sc ii}] (K)& 10\,080 & 9\,660 $\pm$ 400 \\ 
$T_{\rm e}$ [S~{\sc iii}] (K)& 9\,200 & 9\,260 $\pm$ 420 \\ 
$n_{\rm e}$ [S~{\sc ii}] (cm$^{-3}$)& 680 & 640 $\pm$ 200  \\ 
$n_{\rm e}$ [Cl~{\sc iii}] (cm$^{-3}$)& 1\,040 & 720 $\pm$ 480 \\ \hline
\end{tabular}
}
\label{table_te_ne}
\end{table}

In addition, we performed a radial and 2D analysis of the observed (MUSE) and modelled emission line maps. Pseudo-slits of size 1\arcsec $\times$ 30\arcsec were placed towards the north, east, south, and west directions, extending from the central star to the outer parts of NGC~3132. This section presents the results from the analysis of the western side of the nebula, while the other three positions are presented in Appendix~\ref{sat_rad_app}.

In Fig.~\ref{Ne_Te_rad}, we present the radial distribution of $n_{\rm e}$ (upper panel) and $T_{\rm e}$ (lower panel). The blue line with error bars corresponds to the observed MUSE data, while the orange line represents the modelled data. $T_{\rm e}$ are determined using the same temperature diagnostic line ratios for both modelled and MUSE data. Our derived radial distribution of $n_{\rm e}$ from MUSE is in agreement with predictions of the model, starting with a low, nearly constant value of $\sim$ 400~cm$^{-3}$ up to $\sim$ 12\arcsec~from the central star, then increasing to $\sim$ 1\,000 cm$^{-3}$ at the position of the rim at a distance of 18\arcsec, and finally decreasing at larger distances. MUSE radial profiles of $n_{\rm e}$ for all four pseudo-slit positions are in good agreement with the model ones (Fig.~\ref{ne_rad_north}). It should be noted that an offset of $\sim$ 2\arcsec~is observed between the observed and modelled data on the northern side of NGC~3132, likely due to weak observational constrains in that region \citep[see Fig. 3 in][]{hektor2025}. Moreover, there is a bump at a distance of $\sim$~5\arcsec from the central star towards the east direction, which we attribute to a filamentary structure not included in the model.

\begin{figure}
    \centering
    \includegraphics[width=0.45\textwidth]{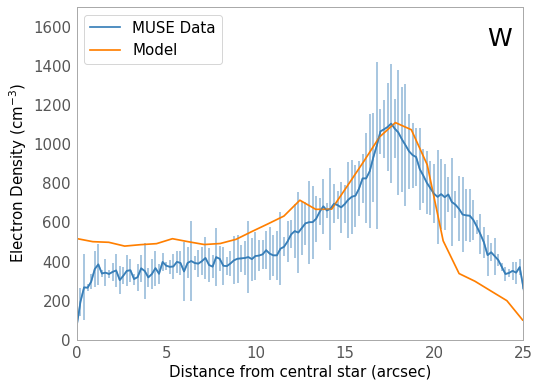}
    \includegraphics[width=0.45\textwidth]{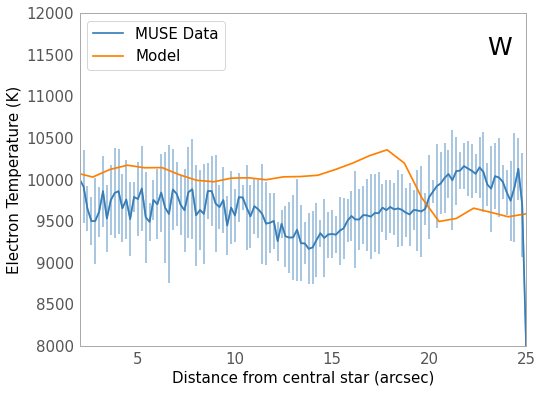}
    \caption{Top panel: radial profile of $n_{\rm e}$ [S~{\sc ii}] west from the CSPN. Bottom panel: radial profile of $T_{\rm e}$ [N~{\sc ii}] west from the CSPN. Blue colour denotes MUSE data, while orange model values.} 
    \label{Ne_Te_rad}
\end{figure}

Regarding $T_{\rm e}$, the model reproduces MUSE's nearly flat distribution throughout the nebula, within the uncertainties. The variations in $T_{\rm e}$ in the observed data, are attributed to the low signal-to-noise ratio of the auroral [N~{\sc ii}]~$\lambda$5755 emission line at different distances. Overall, the difference between the observations and the model is $\sim$~5$\%$, can explain most of the discrepancies in the ionic abundances (see Sect. \ref{sect_muse_model_ions}) and, consequently, in the total elemental abundances.

\begin{figure*}     
    \centering{
    \includegraphics[width=0.79\textwidth]{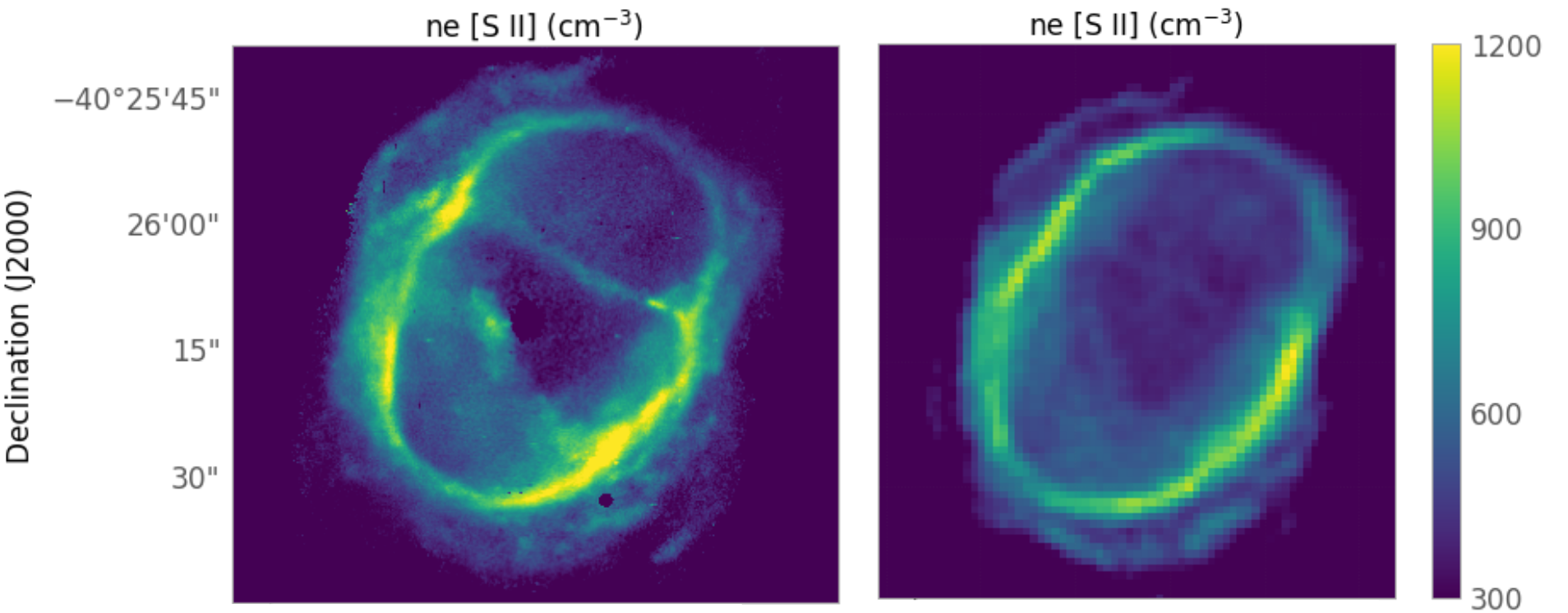}
    \includegraphics[width=0.79\textwidth]{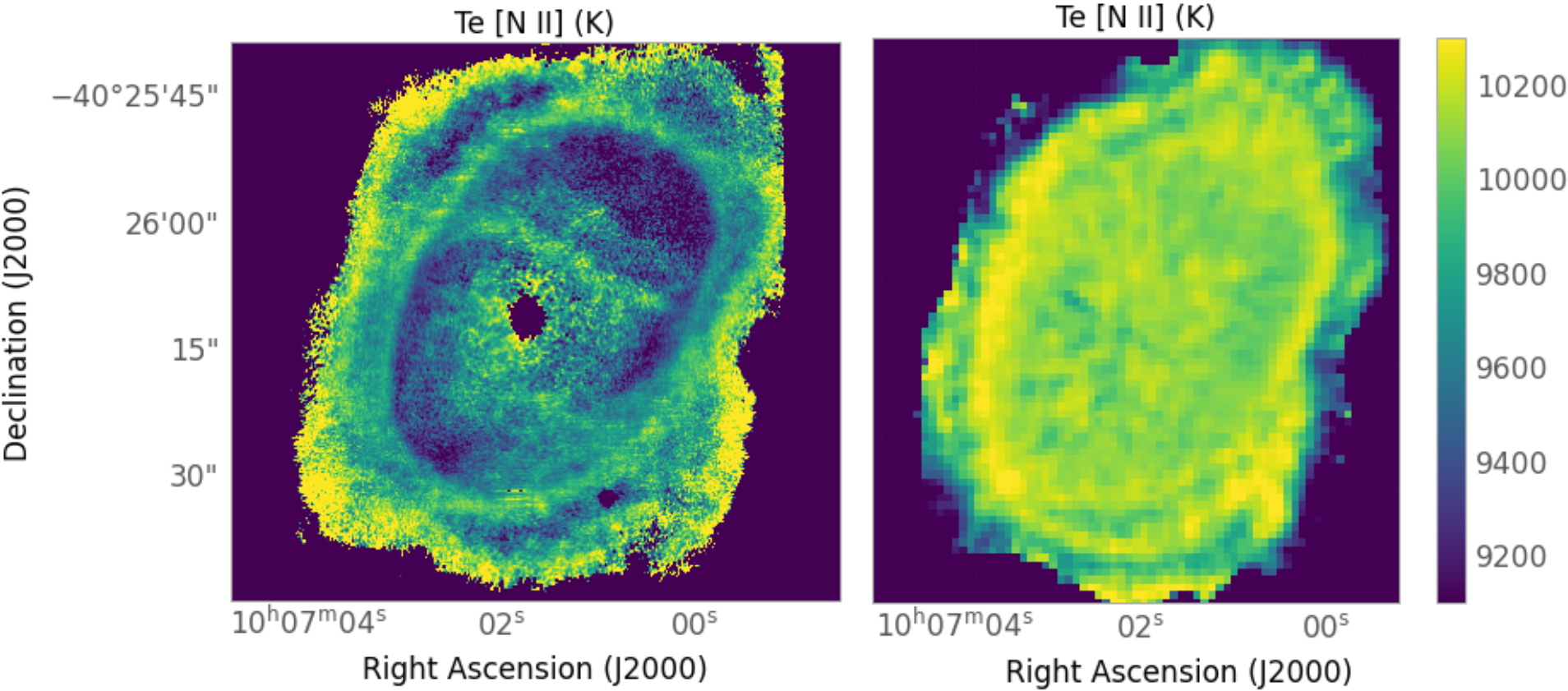}
    }
    \caption{NGC~3132 2D map of $n_{\rm e}$ [S~{\sc ii}] (top panels) and $T_{\rm e}$ [N~{\sc ii}] (bottom panels). MUSE data are located in the left panels, while model data in the right panels.}
    \label{2d_Ne_Te_data}
\end{figure*}

Figure~\ref{2d_Ne_Te_data} exhibits the observed (left panels) and modelled (right panels) 2D maps of $n_{\rm e}$ [S~{\sc ii}] (top panels) and $T_{\rm e}$ [N~{\sc ii}] (bottom panels). For a fair comparison, electron temperature and density are estimated in both the MUSE and modelled data using the same approach--line ratios sensitive to these quantities. Electron density peaks at the nebular rim and in the inner filamentary structure along the east-west direction. The model accurately reproduces the distribution of $n_{\rm e}$ [S~{\sc ii}], which varies from 100~cm$^{-3}$ to 930~cm$^{-3}$, with a median value of 450~cm$^{-3}$. Two inner filaments are only barely seen in the modelled map. The overall spatial distribution of $T_{\rm e}$ [S~{\sc ii}] is approximately reconstructed, though with slightly higher values, ranging from 8\,460 K to 10\,300 K, with a median value of 9\,980~K. 

Additional diagnostic maps, such as $T_{\rm e}$ [S~{\sc iii}] and $n_{\rm e}$ [Cl~{\sc iii}] for the moderately ionised plasma, were also generated. In general, we found that $T_{\rm e}$ [S~{\sc iii}]>$T_{\rm e}$ [N~{\sc ii}], while $n_{\rm e}$ [Cl~{\sc iii}]>$n_{\rm e}$ [S~{\sc ii}], consistent with \citet{Ibero2020}.

The 2D results for $T_{\rm e}$ and $n_{\rm e}$ derived from MUSE data corroborate those reported by \citet{Ibero2020}, with small differences--mainly in $n_{\rm e}$ from [S~{\sc ii}] and [Cl~{\sc iii}]--attributed to different atomic data used in each study. Thus, the {\sc satellite} code effectively reproduces previous findings, making NGC~3132 the fourth successful, example alongside NGC~7009, NGC~6778 \citep{akras2022may}, and NGC~3242 \citet{lydia2025}. Overall, the first spatial comparison of the electron temperature ($T_{\rm e}$) and electron density ($n_{\rm e}$) between the observations and the 3D model shows good agreement, demonstrating the model's ability to accurately reproduce the physical conditions of the ionized gas in NGC~3132. 

\subsection{Ionic and total abundances}
\label{sect_muse_model_ions}

Ionic abundances of He$^+$, He$^{+2}$, O$^+$, O$^{+2}$, N$^+$, S$^+$, S$^{+2}$, Cl$^{+2}$ and Ar$^{+2}$, atomic abundances of O$^0$ and N$^0$ (Figs.~\ref{He_ion_abund}, \ref{O_ion_abund}, \ref{N_S_ion_abund} and \ref{Cl_Ar_ion_abund}) as well as their total elemental abundances (Figs. \ref{he_abund} and \ref{rest_abund}) are also computed for both the observed and modelled emission line maps, using the ICFs provided by \citet{DIMS2014} (hereafter DIMS14), for all pseudo-slits. 

A noticeable variation in ionic abundances is observed across different pseudo-slits. In particular, He$^+$, O$^+$, N$^+$, S$^+$, Ar$^{+2}$ abundances are higher in the outer regions of the nebula (pseudo-slits no. 1, 2, 8 and 9), while He$^{+2}$ and O$^{+2}$ are more prominent in the inner regions (pseudo-slits no. 3, 4, 5, 6 and 7), as illustrated in Figs.~\ref{He_ion_abund}, \ref{O_ion_abund}, \ref{N_S_ion_abund} and \ref{Cl_Ar_ion_abund}. This distribution reflects the expected ionization stratification within the nebula, where high-ionization species are concetrated towards the central regions, and low-ionization species dominate the outer layers. The differences between the model and the MUSE data can be attributed, at least in part, to the afforementioned discrepancies in $T_{\rm e}$ estimates.

In contrast, total elemental abundances are expected to remain relative constant throughout the nebula, since ICFs are designed to account for different ionization stages. However, this is not what we observe in our study. Modelled He, N, S, and Ar abundances, as well as their MUSE counterparts, exhibit higher values in the outer regions of the nebula (pseudo-slits no. 1, 2, 8 and 9) compared to the inner ones (pseudo-slits no. 3, 4, 5, 6 and 7), as shown in Figs.~\ref{he_abund} and \ref{rest_abund}. In addition, the corresponding ICFs are plotted in Fig.~\ref{rest_ICFs}. This discrepancy mainly stems from inaccuracies in the ionic abundances and possible limitations of the ICF formulae. Since ICFs are defined for integrated spectra of entire nebulae, applying them to slits covering specific parts of the ionization stratification (such as pseudo-slits no. 1, 2, 8, 9) can lead to inaccuracies. This results in an apparent enhancement of total elemental abundances in the outer parts of the nebula. Notably, total abundances derived from the inner pseudo-slits are consistent with those obtained from the integrated observed and modelled spectra (pseudo-slit no. 10), indicating that partial coverage of the ionization structure introduces systematic uncertainties in the abundance derivations.

\begin{figure}
    \centering
    \includegraphics[width=0.45\textwidth]{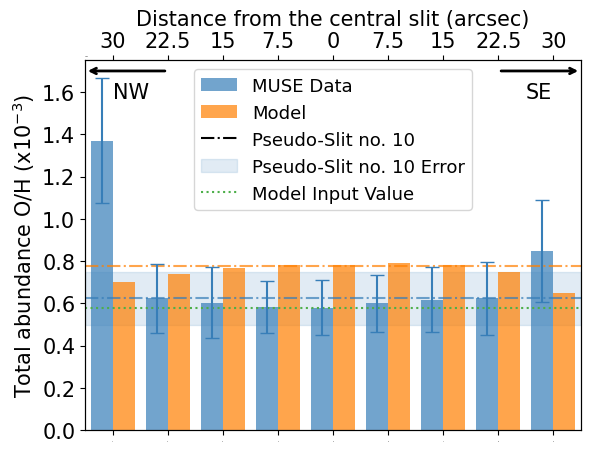}
    \includegraphics[width=0.45\textwidth]{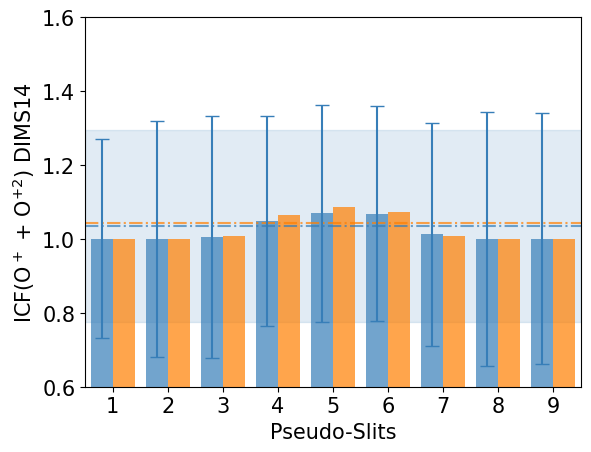}
    \caption{Total Oxygen abundance (top panel) and ICF(O$^+$ + O$^{+2}$) estimates (bottom panel) from DIMS14 as estimated from MUSE data and the 3D model. Blue colour denotes the MUSE data, while orange the model values for each pseudo-slit. The dashed lines show the values for the entire nebula (slit no. 10) and the light blue regions indicate the error of the observed measurement. The dotted green line corresponds to the abundance that was used as input for the model.} 
    \label{model_abund_slits}
\end{figure}

As an illustrative example, we present the O abundance and the corresponding ICF in the upper and lower panel of Fig.~\ref{model_abund_slits}, respectively. Oxygen abundances derived from the MUSE data, show no significant variations among most of the pseudo-slits, except pseudo-slits no. 1 and 9, which cover the outer parts of the nebula and result in incorrect abundance estimates. This can be explained by the very low ICF, resulting from the detection of both O$^+$ and O$^{+2}$ emission lines. In contrast, the oxygen abundances derived from the modelled line maps are slightly higher at the inner regions compared to the outer ones.

For Cl, abundances were computed only for the inner region (pseudo-slits no. 4, 5, 6 and 7). Despite the detection of the [Cl~{\sc iii}]~$\lambda\lambda$5517, 5538 emission lines in the outer regions of the nebula, Cl abundances were not calculated in these regions because the ionisation degree is out of the validity range of the ICF scheme proposed by DIMS14.

\begin{figure}     
    \centering{
    \includegraphics[width=0.42\textwidth]{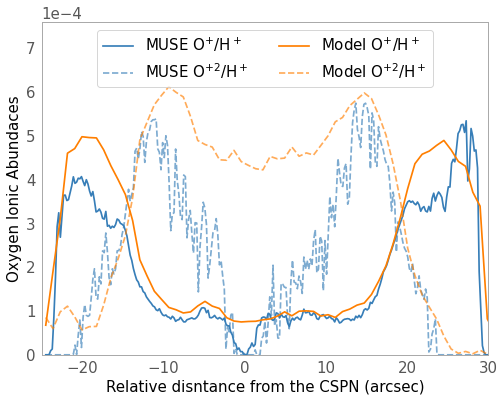}}
    \caption{Radial profiles of singly (solid line) and doubly (dashed line) ionised oxygen abundances from MUSE data (blue lines) and the model (orange lines) across pseudo-slit no. 5.}
    \label{abund_rad}
\end{figure}

Ionic abundance 2D maps were also generated\footnote{2D maps of total abundances were not created since ICFs should not be used in spatially resolved data.}. As expected, higher ionisation species are concentrated in the inner cavity, while lower ionisation species are found around the nebular rim. As an example, Fig.~\ref{abund_rad} presents the radial profiles of singly and doubly ionised oxygen abundances across pseudo-slit no. 5. O$^+$ and O$^{+2}$ abundances displayed in Figure \ref{abund_rad} are estimated from [O~{\sc ii}]$~\lambda$7320 and [O~{\sc iii}]~$\lambda$5007 fluxes, respectively. This corroborates the problem of measuring chemical abundances using slits that do not pass through the central region of nebulae and fail to cover the entire ionisation structure. Moreover, the model reproduces reasonably the observed radial distribution of O$^+$ at smaller radial distances, but it slightly overestimates it near the nebular rim ($\sim~-20$\arcsec~and $\sim~+25$\arcsec). However, O$^{+2}$ is overestimated for radial distances closer to the central star. The differences in $T_{\rm e}$ and $n_{\rm e}$ derived from the model and MUSE data (Table \ref{table_te_ne}) can account for part of the discrepancies shown in Figs.~\ref{He_ion_abund}, \ref{O_ion_abund}, \ref{N_S_ion_abund} and \ref{Cl_Ar_ion_abund}. For example, discrepancies in $T_{\rm e}$ and $n_{\rm e}$ (primarily the former) translates into a $-$24.5$\%$ difference for O$^{+}$ and a +3.0$\%$ difference for O$^{+2}$ in pseudo-slit no. 10. Overall, the comparison between the observed and modelled ionic abundance maps reveals differences that exceed the observational uncertainties.

\subsection{Diagnostic diagrams}

\begin{figure*}     
    \centering{
    \includegraphics[width=0.85\textwidth]{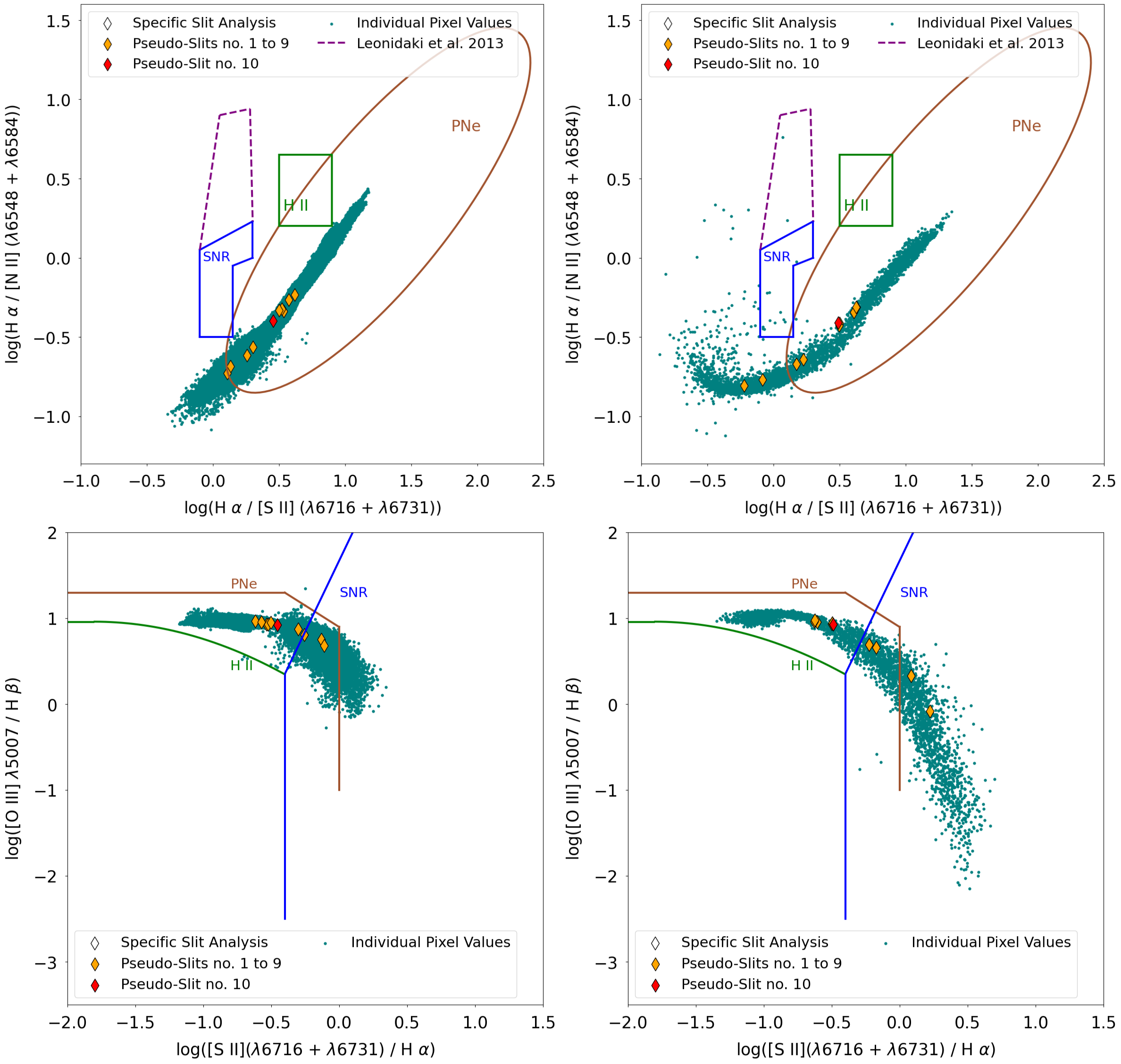}}
    \caption{Emission line diagnostic diagrams for NGC~3132. Left panels correspond to observational data, while right panels show modelled results. For the modelled data, only pixels with flux above the MUSE detection limit ($\sim$10$^{-17}$ erg$\cdot$s$^{-1}$) are included. Cyan dots represent the spaxel values, and diamonds correspond to specific slit values, with yellow denoting pseudo-slits no. 1-9 and red diamond indicating pseudo-slit no. 10 (whole nebula). The loci of PNe, SNRs and H~{\sc ii} regions are also depicted with brown, blue and green colour, respectively \citep[e.g.][]{diagnostics}. The magenta dashed-line area indicates the position of SNRs in lower metallicity galaxies \citep{leonidaki2013}.}
    \label{diagnostic_data}
\end{figure*}

We also examined the distribution of specific line ratios commonly used to distinguish different astrophysical objects and excitation mechanisms \cite[e.g.,][]{Sabbadin1977,bpt1981}. Figure~\ref{diagnostic_data} illustrates the line ratios log(H~$\upalpha$/[N~{\sc ii}]) versus log(H~$\upalpha$/[S~{\sc ii}]) (top panels) and log([O~{\sc iii}]/H~$\upbeta$) versus log([S~{\sc ii}]/H~$\upalpha$) (bottom panels) for both the observed (left panels) and modelled (right panels) line maps. The cyan dots represent the emission line ratio for individual spaxels from the 2D map \footnote{Only model pixels with values above the MUSE detection limit ($\sim$10$^{-17}$ erg$\cdot$s$^{-1}$) are considered.}, diamonds represent the ratios from the pseudo-slits (yellow for slits no. 1-9 and red for slit no. 10) in the specific slit module of {\sc satellite}. The location of PNe, SNRs, and H~{\sc ii} regions are also delineated in brown, blue, and green, respectively. 

Concerning the integrated values (diamond shape), which correspond to the pseudo-slits that cover the central regions (pseudo-slits no. 4, 5, 6, 7) and the whole nebula (pseudo-slit no. 10 in Fig.~\ref{SAT_spec}), they lie well within the PNe region regime in all four panels of Fig.~\ref{diagnostic_data}. However, the corresponding yellow diamonds for the pseudo-slits that cover the outer regions of the nebula (pseudo-slits no. 1, 2, 8, 9), are located near the boundaries of the PNe region or in the SNR regime. In particular, these four pseudo-slits are characterized by significantly more intense emission from low-ionisation lines relative to the central nebula, as expected due to the ionisation stratification. This explains why these pseudo-slits result in higher He, N, S, Ar abundances compared to the ones derived from pseudo-slits covering the central region of the nebula (Figs \ref{he_abund} and \ref{rest_abund}). 

The interpretation of individual spaxel values in the diagnostic diagrams based on integrated observations carries inherent risks, and conclusions drawn from such diagrams should be approached with caution \citep{morisset2018}. Additionally, high(low)-ionisation ratios are not a priori an indication of shock-heated gas. However, it is evident that individual spaxels (cyan dots) follow the overall trend provided by the integrated values (yellow diamonds) in Fig.~\ref{diagnostic_data}. This consistency justifies the inclusion of individual spaxel values in the diagrams. Thus, it is not appropriate to perform a spaxel-by-spaxel analysis to determine their excitation mechanism, but the overall distribution of the spaxels provides valuable insights. The distribution of the modelled line ratios is similar to those from the observation, ranging within the same values, except for a few spaxels with low H~$\upalpha$/[N~{\sc ii}], H~$\upalpha$/[S~{\sc ii}], and [O~{\sc iii}]/H~$\upbeta$ line ratios. These spaxels are distributed in the outer part of the nebula, where the low-ionisation gas enhances the emission of low-ionisation species and places them in shock-heated gas regime \citep[e.g.][]{kopsaxili2020}. It is worth noting that a pure photoionization model succesfully reproduces observational properties of NGC~3132, without the need of shock excitation. The same conclusion has been drawn for the VIMOS@VLT IFU data of the PN Abell 14 \citep{akras2020}.

\section{Comparing the empirical and photoionisation modelling approaches}
\label{sec4}

Leveraging the recent 3D photoionization model of \citet{hektor2025}, we evaluate the accuracy of the commonly used empirical method when applied to simulated observations and assess how well the model's input parameters can be recovered. This approach allows us to gauge where the model and empirical methods agree, as well as to explore their limitations under different observational conditions. We carry out two distinct tests: (1) comparing the results of the model and empirical methods under the same set of observational configurations; and (2) evaluating how accurately the empirical method recovers the model values when the latter are assumed to represent the truth. Understanding these aspects is crucial for evaluating the level of detail that can be achieved in studies of mass-loss processes in PNe using different methodologies.

\subsection{Evaluation of the empirical method for estimating elemental abundances}

With the physical and chemical structure of the nebula fully defined in the 3D model \citep{hektor2025}, we can evaluate the accuracy of the empirical method specifically in deriving the total elemental abundances. By applying the empirical method to modelled emission lines maps or integrated fluxes (i.e., synthetic data), we eliminate observational uncertainties, allowing us to isolate the intrinsic limitations of the method itself.

To investigate this, we examined the discrepancy in the oxygen abundances derived from pseudo-slits no. 5 and 10 and the model input abundance (Fig.~\ref{model_abund_slits}). These pseudo-slits were specifically chosen because they encompass all ionisation zones. This is similar to the approach used in 1D photoionisation models, from which the ICF formulae are derived \citep{DIMS2014}.

In a real planetary nebula, the distribution of gas and dust, along with its corresponding ionization structure, can be complex, leading to a wide range of observed morphologies. As a result, observations using long-slits may sample distinct regions and combinations of ionization zones projected along the line of sight, depending on where the slit is placed. Since ICF formulae available in the literature are constructed from one-dimensional spherical models --assuming integrated line fluxes that include all ionizations zones-- they should not be applied to slit positions that fail to sample the full ionization structure.

\begin{figure}     
    \centering{
    \includegraphics[width=0.41\textwidth]{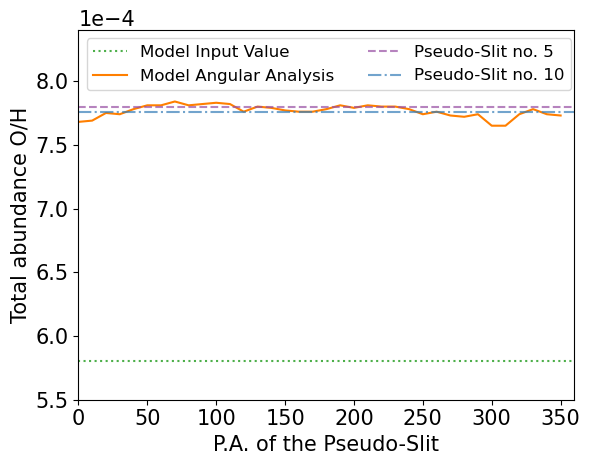}}
    \caption{3D model total oxygen abundance as estimated by {\sc satellite} angular module (solid orange line), by specific slit no. 5 (dashed purple line) and no. 10 (dashed-dot blue line). The dotted green line corresponds to the abundance that was used as input for the model.}
    \label{abund_o_ang}
\end{figure}

To illustrate the importance of including all ionizations zones in slit-based observations, we derived abundances for multiple slit orientations (ranging from 0$^{\circ}$ to 360$^{\circ}$), all passing through the center of the nebula. In Fig.~\ref{abund_o_ang}, we present the total oxygen abundancies obtained using the {\sc satellite} code for each orientation, which reveal no variation with position angle. The obtained values are also consistent to those derived from pseudo-slits no. 5 and no. 10, with similar results found for the other elements. Even though the slit orientation had no impact on the derived abundance, none of the configuratios succeeded in recovering the adopted input abundance of the model.

Current ICFs formulae produce different abundance estimates and tend to overestimate the chemical abundances by a notable amount, ranging from 8$\%$ to 35$\%$. The input abundances for the model are listed in the second column of Table~\ref{table_comp}. The remaining columns present the chemical abundances derived from different approaches. Specifically, the third column lists the abundances obtained using the method of DIMS14 for pseudo-slits no. 5 and 10. The difference between the model input values and these two estimates is given as a percentage in the last two columns. Considering that both He ions are observed and no ICF is necessary, the total He abundance derived from the emission line maps is the least overestimated relative to the input value, compared to the other elements. Therefore, the observed discrepancies between the model input abundances and those estimated by the empirical method may be influenced by two factors: projection effects and/or limitations in the accuracy of ICF formulae, at least in the case of NGC~3132.

\begin{table}
\centering
\caption{Elemental abundances estimated from pseudo-slits (PS) no. 5 and 10, covering the central part and the whole nebula, respectively. The fourth column lists the percentage difference between the model input and the results from the empirical method. Total abundances of N, O, S, Cl and Ar are estimated using ICF formulae from DIMS14.}
\resizebox{0.5\textwidth}{!}{%
\begin{tabular}{|c|cc|cc|cc|}
\hline
Abundances & \multicolumn{2}{c|}{Model Input} & \multicolumn{2}{c|}{Model}                & \multicolumn{2}{c|}{Percentage (\%)}      \\ 
           & \multicolumn{1}{c|}{}     &      & \multicolumn{1}{c|}{PS no. 5} & PS no. 10 & \multicolumn{1}{c|}{PS no. 5} & PS no. 10 \\ \hline
He/H $\times$ 10$^{-1}$ & \multicolumn{2}{c|}{1.17} & \multicolumn{1}{c|}{1.26} & 1.29 & \multicolumn{1}{c|}{+7.7}  & +10.2 \\ 
N/H $\times$ 10$^{-4}$  & \multicolumn{2}{c|}{2.27} & \multicolumn{1}{c|}{2.65} & 2.86 & \multicolumn{1}{c|}{+16.7} & +26.0 \\ 
O/H $\times$ 10$^{-4}$  & \multicolumn{2}{c|}{5.80} & \multicolumn{1}{c|}{7.80} & 7.76 & \multicolumn{1}{c|}{+34.5} & +33.8 \\ 
S/H $\times$ 10$^{-5}$  & \multicolumn{2}{c|}{1.06} & \multicolumn{1}{c|}{1.34} & 1.39 & \multicolumn{1}{c|}{+26.4} & +31.1 \\ 
Cl/H $\times$ 10$^{-7}$ & \multicolumn{2}{c|}{1.70} & \multicolumn{1}{c|}{2.10} & 2.03 & \multicolumn{1}{c|}{+23.5} & +19.4 \\ 
Ar/H $\times$ 10$^{-6}$ & \multicolumn{2}{c|}{3.10} & \multicolumn{1}{c|}{2.82} & 2.96 & \multicolumn{1}{c|}{-9.0} & -4.5 \\ 
N/O                    & \multicolumn{2}{c|}{0.39} & \multicolumn{1}{c|}{0.34} & 0.37 & \multicolumn{1}{c|}{-12.8} & -5.1 \\ \hline
\end{tabular}
}
\label{table_comp}
\end{table}

\subsection{The case of oxygen}

\begin{figure}
    \centering
    \includegraphics[width=0.45\textwidth]{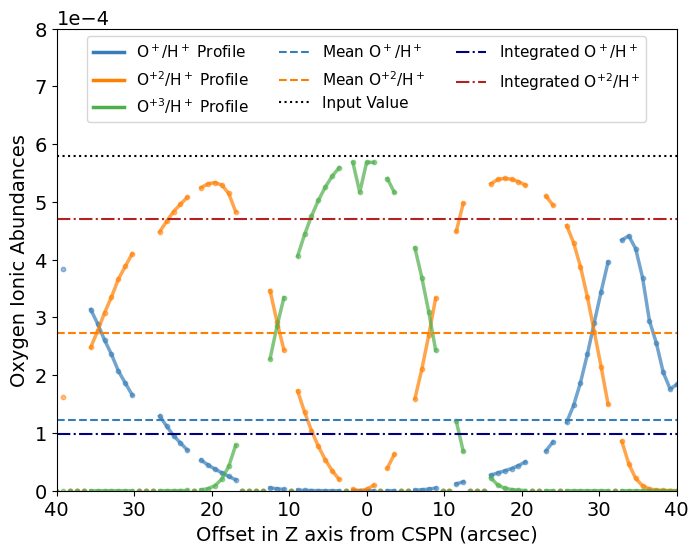}
    \caption{O$^+$ (blue solid lines), O$^{+2}$ (orange solid lines) and O$^{+3}$ (green solid lines) abundances profiles along the line of sight for the central spaxel of the NGC~3132 model. Dashed lines correspond to the mean of O$^+$ (blue) and O$^{+2}$ (orange) distributions. Dashed dotted lines correspond to O$^+$ (dark blue) and O$^{+2}$ (dark orange) abundances estimated from already integrated along the lines of sight fluxes. The black dotted line corresponds to model input value and zero values in the profiles are due to filling factor adopted in the model.} 
    \label{3d_2d_1pix}
\end{figure}

We further examined the impact of projection effects on the elemental abundances in NGC~3132. Since the model is constructed in 3D space, its input abundances correspond to the full spatial structure, as well. However, when integrating the model along a specific line of sight to produce the 2D emission maps, some information is inevitably lost, depending on the distribution along the line of sight. Notably, we noticed that only for oxygen, the sum of O$^+$ and O$^{+2}$ ionic abundances in pseudo-slits no. 5 and 10 exceed the input O abundance of the 3D model. To verify the consistency of the model, we summed the modelled 3D cubes of all oxygen ionization stages and confirmed that the total O abundances matches the input value, as expected. This proves that the apparent overestimation of O$^+$ and O$^{+2}$ from the 2D maps is due to the integration over the line of sight. Projection effects directly impact flux measurements and the $T_{\rm e}$ estimates\footnote{$n_{\rm e}$ has less impact on the ionic abundance estimates.}, which in turn influence the derived ionic abundances. To further demonstrate this effect, Fig.~\ref{3d_2d_1pix} presents the O$^+$ (blue), O$^{+2}$ (orange) and O$^{+3}$ (green) abundances profiles along the line of sight of a single spaxel (the zero values in the profiles are due to filling factor adopted in the model). The mean values of these profiles are marked by blue and orange dashed lines, respectively. In contrast, if the integrated fluxes of [O~{\sc ii}] and [O~{\sc iii}] emission maps for the same spaxel were used, the resulting ionic abundances for O$^+$ (dark blue dot-dashed line) and O$^{+2}$ (dark orange dot-dashed line) would differ.

\begin{figure}
    \centering
    \includegraphics[width=0.55\textwidth]{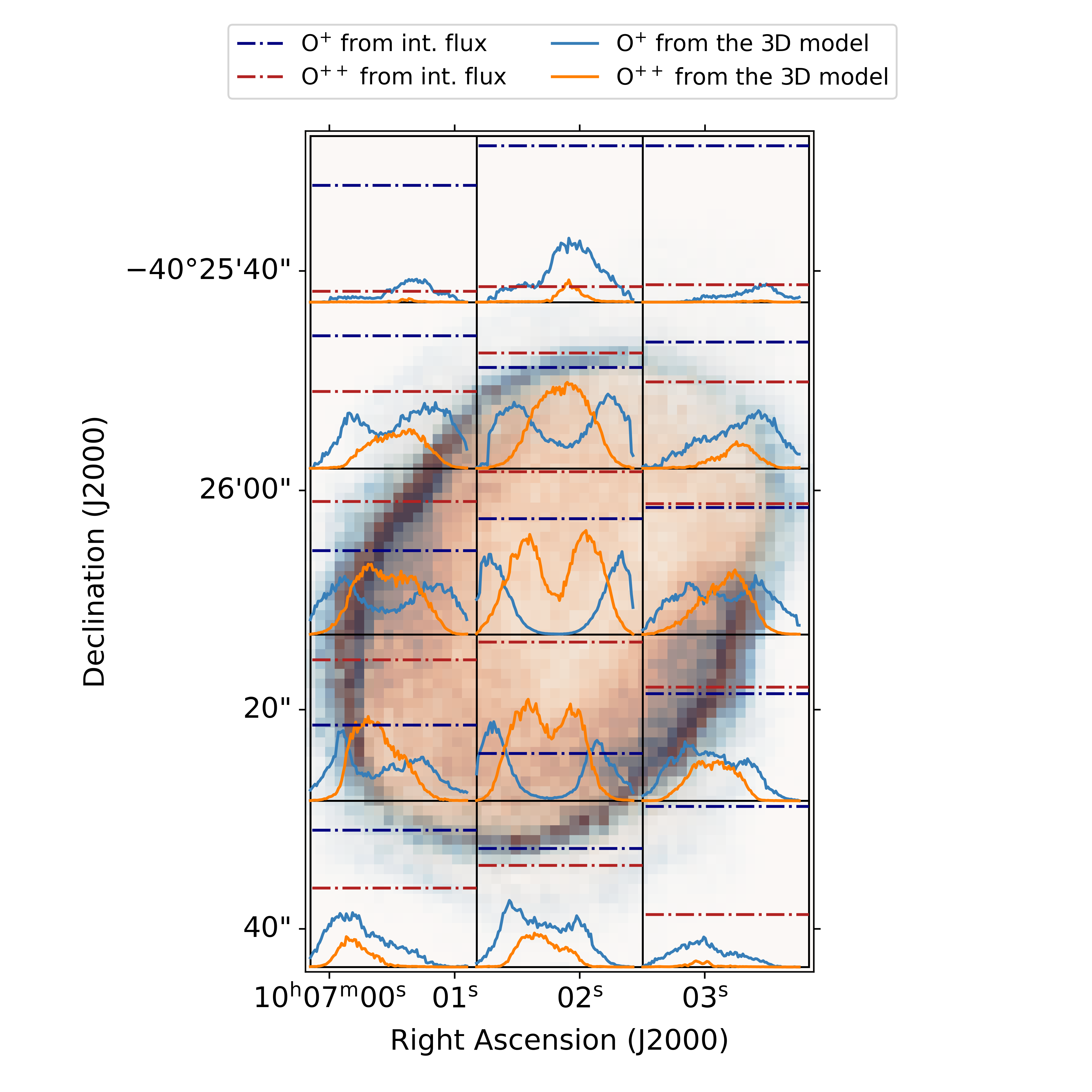}
    \caption{O$^+$ (blue solid) and O$^{+2}$ (orange solid) abundances profiles along the line of sight for 15 selected regions in NGC~3132. Dashed lines in each subplot represent O$^+$ (dark blue) and O$^{+2}$ (dark orange) abundances estimated from already integrated along the lines of sight fluxes for each region. The upper y-value for each subplot corresponds to the model input value. The background image combines [O~{\sc ii}]~$\lambda$7320 (blue) and [O~{\sc iii}]~$\lambda$5007 (orange) emission maps.} 
    \label{3d_2d_comp}
\end{figure}

Furthermore, in order to show that the same effect is independent of the position selected in the nebula, we created Fig.~\ref{3d_2d_comp}, which includes 15 regions covering the entire nebula. We show the average O$^+$ (blue lines) and O$^{+2}$ (orange lines) abundances profiles along the line of sight for each region. Simultaneously, the ionic abundances derived from the already integrated fluxes along the line of sight are displayed in the same regions, with dark blue (O$^+$) and dark orange (O$^{+2}$) dot-dashed lines. The background of Fig.~\ref{3d_2d_comp} combines [O~{\sc ii}] $\lambda$7320 and [O~{\sc iii}] $\lambda$5007 emission maps in blue and orange, respectively.

Notably, in all regions, except the top left one, the sum of the O$^+$ and O$^{+2}$ abundances estimated from the integrated fluxes exceeds the input value of the 3D model, similar to the cases of pseudo-slits no. 5 and 10. Additionally, the first and fifth row regions (from the bottom) exhibit significant overestimation in the integrated O$^+$ abundance. In contrast, in the central regions of the nebula (second and third row regions) the integrated O$^{+2}$ abundance is overestimated. Consequently, the projection effect has a non-negligible impact on the estimations of various ionic abundances in distinct regions of the nebula.

Specifically for the case of oxygen, we assessed the impact of projection effects on $T_{\rm e}$ (derived from [N~{\sc ii}] and [S~{\sc iii}] line ratios) and $n_{\rm e}$ (derived from [S~{\sc ii}] and [Cl~{\sc iii}] line ratios), which are used in the computation of O$^+$ and O$^{+2}$ abundances. Then, we translated the individual differences in $T_{\rm e}$ and $n_{\rm e}$ into discrepancies in the O$^+$ and O$^{+2}$ ionic abundances (Table \ref{projection}). Singly ionized oxygen is underestimated in both pseudo-slits no. 5 (-6.4$\%$) and no. 10 (-24.0$\%$), while doubly ionized oxygen is overestimated in both slits (+18.5$\%$ and +5.5$\%$, respectively).

\begin{table}
\centering
\caption{$T_{\rm e}$, $n_{\rm e}$ and O ionic abundances estimated for pseudo-slits (PS) no. 5 and 10. The "True" column shows the values directly extracted from the 3D model (unaffected by projection), while the "Int." (i.e. integrated) column lists the corresponding values derived from simulated observations.}
\resizebox{0.5\textwidth}{!}{%
\begin{tabular}{|c|cc|cc|}
\hline
\multirow{3}{*}{Projection Effect} & \multicolumn{2}{c|}{PS no. 5}    & \multicolumn{2}{c|}{PS no. 10} \\  
                                   & \multicolumn{1}{c|}{True} & Int. & \multicolumn{1}{c|}{True} & Int. \\ \hline
{$T_{\rm e}$ ([N~{\sc ii}]) (K)}       & \multicolumn{1}{c|}{10018} & 10116 & \multicolumn{1}{c|}{9751} & 10081  \\
{$n_{\rm e}$ ([S~{\sc ii}]) (cm$^{-3}$)}       & \multicolumn{1}{c|}{560}  & 797  & \multicolumn{1}{c|}{398}   & 682  \\
{O$^+$ / H$^+$ ($\times$ 10$^{-4}$)} & \multicolumn{1}{c|}{2.66}  & 2.49 & \multicolumn{1}{c|}{4.08}   & 3.10 \\
{$T_{\rm e}$ ([S~{\sc iii}]) (K)}      & \multicolumn{1}{c|}{9886} & 9188 & \multicolumn{1}{c|}{9703}  & 9206 \\
{$n_{\rm e}$ ([Cl~{\sc iii}]) (cm$^{-3}$)}     & \multicolumn{1}{c|}{786}  & 1165 & \multicolumn{1}{c|}{516}   & 1042 \\
{O$^{+2}$ / H$^+$ ($\times$ 10$^{-4}$)}  & \multicolumn{1}{c|}{3.61}  & 4.28 & \multicolumn{1}{c|}{3.78}   & 3.95\\
\hline
\end{tabular}
}
\label{projection}
\end{table}

It is important to note that different approaches must be used for fluxes and for physical parameters such as $T_{\rm e}$, $n_{\rm e}$ and abundances. Flux is an extensive quantity, and should therefore be integrated (i.e. summed) along the z-axis. In contrast, $T_{\rm e}$, $n_{\rm e}$ are intensive quantities and require a weighted average along the third axis. For temperature or density, the weighted average is typically based on the flux of a relevant diagnostic line. In this study, [N~{\sc ii}] $\uplambda$5755 and [S~{\sc iii}] $\uplambda$9069 were used for $T_{\rm e}$, and [S~{\sc ii}] $\uplambda$6716 and [Cl~{\sc iii}] $\uplambda$5517 were used for $n_{\rm e}$.

Table \ref{projection} shows that plasma diagnostics for intermediate- to high-ionization plasma (i.e. $T_{\rm e}$ and $n_{\rm e}$ derived from [S~{\sc iii}] and [Cl~{\sc iii}] emission line ratios) seem to be more affected by projection effects than those for low-ionization plasma diagnostics. In contrast, the impact of projection on $T_{\rm e}$ and $n_{\rm e}$ to the abundance estimates of singly and doubly ionized oxygen is similar: O$^+$ is underestimated, while O$^{+2}$ is overestimated relative to the abundance derived with the "True" $T_{\rm e}$ and $n_{\rm e}$. These differences lead to an overestimation of 8$\%$ in the total O abundance for pseudo-slit no. 5 and an underestimation of 10$\%$ for pseudo-slit no. 10. The remaining discrepancies between the estimated abundances and the model input values (see fourth column in Table \ref{table_comp}) are due to projection effect acting directly on the flux measurements.

\begin{figure}
    \centering
    \includegraphics[width=0.45\textwidth]{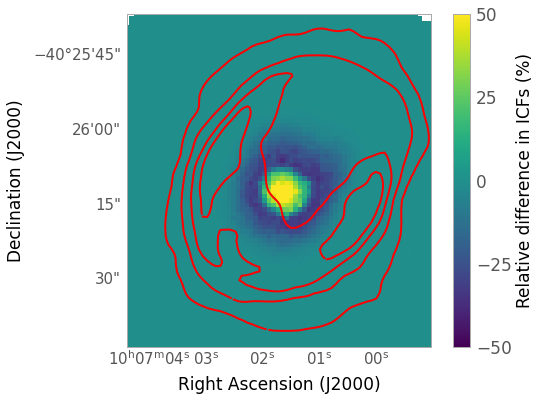}
    \caption{2D map of the relative difference in ICF(O$^+$ + O$^{+2}$) derived from the 3D model and the one using the formulae of DIMS14. The red contours indicate the surface brightness of the H $\upbeta$ emission line.}
    \label{2d_icf}
\end{figure}

Another possible contributor to abundance discrepancies lies in the currently available ICF formulae \citep{DIMS2014}. With access to the full 3D cubes of O$^+$, O$^{+2}$ and O$^{+3}$ abundances from the model, we are able to construct our ICF(O$^+$ + O$^{+2}$) cube that accounts for the contribution of the unobserved O$^{+3}$ ion. Figure~\ref{2d_icf} shows a comparison between the "True" ICF map, derived from the 3D model by averaging along the line of sight, and the corresponding one obtained using DIMS14 formulae. As previously mentioned, current ICFs are not intended for use on spatially resolved data, however it is still worthwhile to assess whether --and to what extent-- they introduce inaccuracies.

Figure~\ref{2d_icf} shows that the "True" ICF derived from the 3D model is considerably higher than that of DIMS14 within a spherical region of $\sim$5$\arcsec$ around the nebular centre, even by as much as 95$\%$ in the central spaxel. In contrast, in the surrounding annular region ($\sim$5$\arcsec$ to $\sim$11$\arcsec$), the "True" ICF is up to 30$\%$ lower than the DIMS14 value. A similar pattern is observed for the corresponding nitrogen plot (upper panel Fig. \ref{rest_2d_ICFs}): within $\sim$15$\arcsec$ of the centre, the "True" ICF is consistently higher than the DIMS14 value, while near the nebular rim it becomes significantly lower. At even larger radial distances, "True" ICF remains slightly smaller than the analytical estimate.

Regarding sulfur (middle panel Fig. \ref{rest_2d_ICFs}), the "True ICF" is substantially higher than that of DIMS14 throughout the whole nebula, with a median excess of $\sim$90$\%$. Similarly, for chlorine (lower panel Fig. \ref{rest_2d_ICFs}), the "True" ICF is always higher by 20$\%$ within a circular region of $\sim$12$\arcsec$ centred northwest from the nebular centre, and by up to 95$\%$ elsewhere. It is important to note that the aforementioned 2D ICF maps are not used for any estimate in this study. To determine ICFs for specific pseudo-slits, we first calculate the corresponding ionic abundances and then derive the ICFs accordingly.

Lastly, we quantified the difference between the two ICFs for pseudo-slits no. 5 and 10, as shown in Table~\ref{icf_table}. To do this, we first find the average O$^+$, O$^{+2}$ and O$^{+3}$ distributions across the spaxels within each pseudo-slit and then we computed the mean of that distribution. Interestingly, we find that the contribution of the ICF(O$^+$ + O$^{+2}$) to the discrepancy in oxygen abundance --derived from integrated fluxes (i.e. simulated observations)--- is zero, leaving the projection effect as the only contributor to the differences. We have to note that $T_{\rm e}$ and $n_{\rm e}$ values derived from emission line ratios were used for these estimations.

\begin{table}
\centering
\caption{Oxygen total abundances and ICF(O$^+$ + O$^{+2}$) estimates (values in parenthesis) derived directly from the 3D model ("True Abund.") and from integrated fluxes ("Empirical Method") for pseudo-slits (PS) no. 5 and 10 using the formulae from DIMS14.}
\resizebox{0.48\textwidth}{!}{%
\begin{tabular}{|c|c|c|c|}
\hline
Position                   & ICF         & True Abund. & Empirical Abund. \\ 
                           &             & $\times$ 10$^{-4}$ & $\times$ 10$^{-4}$ \\ \hline
\multirow{2}{*}{PS no. 5}  & True ICF    & 4.27 (1.16) & 7.80 (1.09) \\ 
                           & DIMS14 ICF  & 4.23 (1.15) & 7.80 (1.09) \\ 
\multirow{2}{*}{PS no. 10} & True ICF    & 2.86 (1.03) & 7.76 (1.04) \\ 
                           & DIMS14 ICF  & 2.86 (1.03) & 7.76 (1.04) \\
\hline
\end{tabular}
}
\label{icf_table}
\end{table}

Therefore, at least in the case of oxygen in NGC~3132, the current ICF formulae perform reasonably well in accounting for the contribution of unobserved ions, for slits covering the whole ionization stratification. This is largely because the spatial distributions of the unobserved O$^{+3}$ ion, as well as He$^+$ and He$^{+2}$ (used in DIMS14 formulae), exhibit relatively roughly spherical structures near the nebular centre, despite the overall morphological complexity of the nebula, similar to the assumptions of the 1D models upon which the ICFs were based on.   

A noticable difference is observed between the "True Abund." results for pseudo-slit no. 5 and 10 (Table \ref{icf_table}). This discrepancy arises from the projection effect on flux estimates, and its behaviour along different slits is difficult to predict a priori. The same effect explains why, despite that there is a good match between the abundance estimates derived from the 3D model ("True Abund.") and integrated fluxes ("Empirical Abund."), the resulting abundances are still significantly higher than the input value of the 3D model (5.8 $\times$ 10$^{-4}$).

\section{Radial analysis of MUSE, \textit{JWST} and \textit{Spitzer} data}
\label{JWST_results}
\subsection{Stratification of ionic, atomic and molecular species}
\label{ionic_strat}

Radial profiles of MUSE line maps and \textit{JWST} NIRCam images were meticulously analysed to examine the ionisation stratification of the nebula. An offset of 0.06\arcsec~between MUSE and \textit{JWST} data must be considered, which accounts for the expansion of the nebula between the two observations\footnote{Given that the MUSE and \textit{JWST} data were obtained in different epochs, the expansion of the nebula should also be taken into account. Considering an expansion velocity of 25 km~s$^{-1}$ for NGC~3132 \citep[chapter 1.6 in][]{osterbrock,guerrero2020} and a distance of $\sim$754 pc \citep{BaylerJones2021}, the expansion of the nebula between the epochs of the MUSE and \textit{JWST} observations was determined to be 0.06\arcsec~or $\sim$2.1$\times10^{-4}$ pc.} The overlap in the far-red spectral range between the MUSE and NIRCam, particularly the [S~{\sc iii}]~$\lambda$9069 line, facilitated the alignment of images from both instruments, enabling the examination of radial profiles for various other lines. It is important to note that the seeing during the observations was approximately 0.7\arcsec~for MUSE, and the FWHM ranged from 0.062\arcsec~and 0.126\arcsec~for different NIRCam filters and between 0.3\arcsec~and 0.6\arcsec~for various MIRI filters (for details, see Sec. \ref{observations}).

Table~\ref{JWST_MUSE_e} lists the peaks of various emission lines for the eastern side of the nebula, in order of their proximity to the central star. An analogous behaviour is also visible in the other directions (see Appendix~\ref{jwst_muse_app}). While the exact peak values may be affected by the FWHM differences between instruments, possibly explaining unexpected variations between species with different ionisation potentials. However, the overall nebula's stratification remains evident in the provided tables.

\begin{table}
\centering
\caption{Radial profiles' peaks of MUSE and \textit{JWST} emissions east from the CSPN.}
\resizebox{0.38\textwidth}{!}{%
\begin{tabular}{|c|c|c|} 
 \hline
 Emission line & 1$^{\rm st}$ peak & 2$^{\rm nd}$ peak \\ [0.2ex] 
 
  & (\arcsec) & (\arcsec) \\
 \hline
 [O~{\sc iii}] 5007~\AA & 16.0 $\pm$ 0.2 & - \\ 
 
 H~$\upalpha$ 6563~\AA & 16.6 $\pm$ 0.2 & - \\ 
 
 H~$\upbeta$ 4861~\AA & 16.6 $\pm$ 0.2 & - \\
 
 [S~{\sc iii}] 18 $\upmu$m & 16.75 $\pm$ 0.11 & - \\
 
 [S~{\sc iii}] 9069~\AA & 16.8 $\pm$ 0.2 & - \\
 
 Pa~$\upalpha$ & 16.84 $\pm$ 0.03 & 22.00 $\pm$ 0.03 \\
 
 [S~{\sc iii}] 9069~\AA~(\textit{JWST}) & 16.94 $\pm$ 0.03 & 22.00 $\pm$ 0.03 \\
 
 Br~$\upalpha$ & 16.95 $\pm$ 0.06 & 22.10 $\pm$ 0.06 \\
 
 H$_{2}$ 2.12 $\upmu$m & 17.26 $\pm$ 0.03 & 22.23 $\pm$ 0.03 \\
 
 [Ne~{\sc ii}] 12.8 $\upmu$m & 17.30 $\pm$ 0.11 & 22.40 $\pm$ 0.11 \\
 
 [N~{\sc ii}] 5755~\AA & 17.4 $\pm$ 0.2 & 21.6 $\pm$ 0.2 \\ 
 
 [N~{\sc ii}] 6584~\AA & 17.4 $\pm$ 0.2 & 21.6 $\pm$ 0.2 \\
 
 H$_{2}$ 3.56 $\upmu$m & 17.45 $\pm$ 0.06 & 22.18 $\pm$ 0.06 \\
 
 H$_{2}$ 4.70 $\upmu$m & 17.51 $\pm$ 0.06 & 22.24 $\pm$ 0.06 \\
  
 PAHs 11.3 $\upmu$m & 17.52 $\pm$ 0.11 & 22.29 $\pm$ 0.11 \\
 
 H$_{2}$ 7.7 $\upmu$m & 17.52 $\pm$ 0.11 & 22.73 $\pm$ 0.11 \\
 
 [S~{\sc ii}] 6731~\AA & 17.6 $\pm$ 0.2 & 21.8 $\pm$ 0.2 \\
 
 [N~{\sc i}] 5199~\AA & 17.8 $\pm$ 0.2 & 22.4 $\pm$ 0.2 \\
 
 [O~{\sc i}] 6300~\AA & 17.8 $\pm$ 0.2 & 22.4 $\pm$ 0.2 \\
\hline 
\end{tabular}}
\label{JWST_MUSE_e}
\end{table}

\begin{figure}   
    \centering{
    \includegraphics[width=0.31\textwidth]{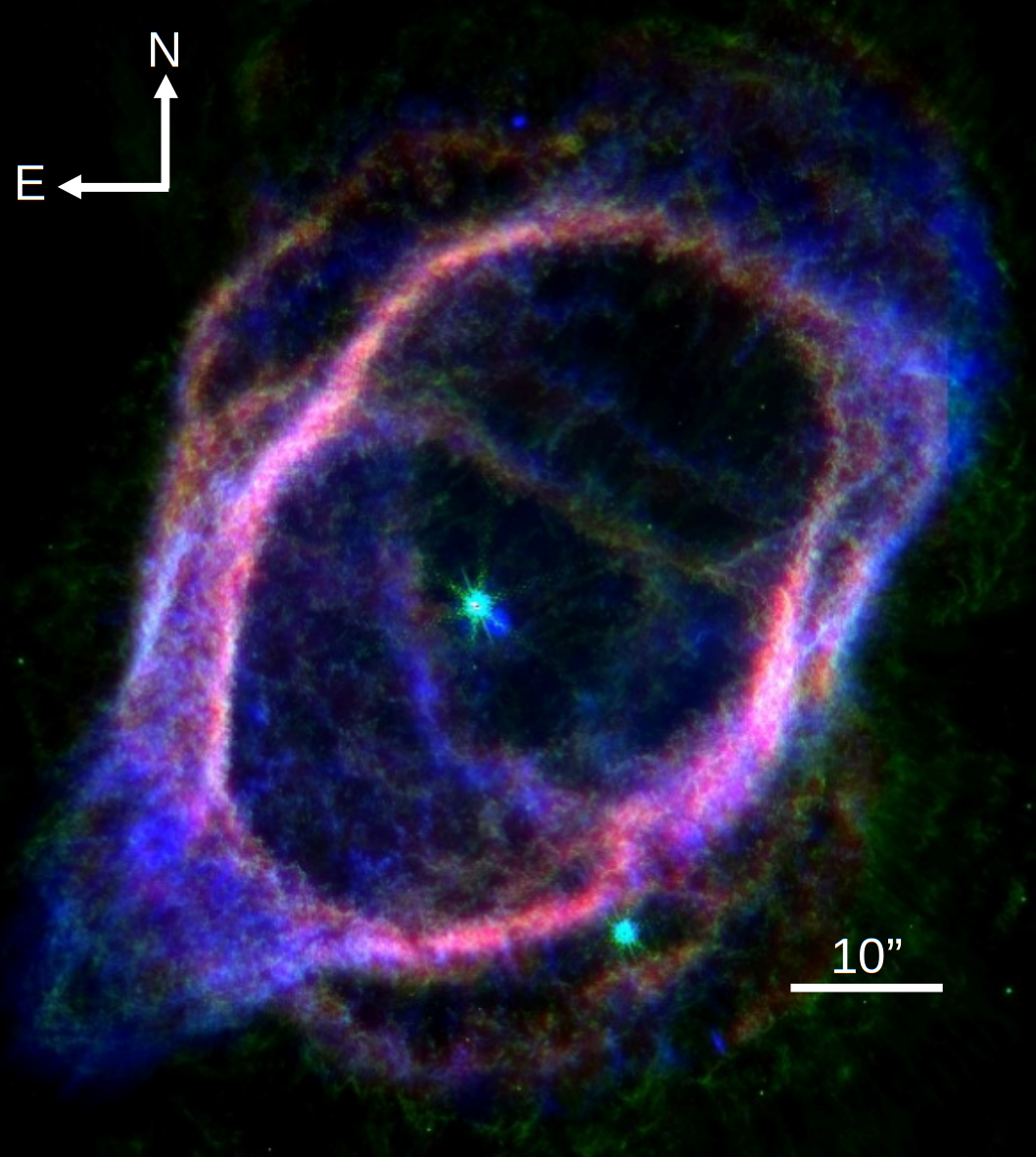}}
    \caption{Composite image of NGC~3132. Red: MUSE [O~{\sc i}] $\lambda$6300, Green: NIRCam@\textit{JWST} H$_2$ 2.12~$\upmu$m, Blue: MIRI@\textit{JWST} PAHs 11.3~$\upmu$m).}
    \label{rgb_radial}
\end{figure}

This analysis verifies the line stratification in the planetary nebula NGC~3132. High to moderate ionisation lines show a peak closer to the central star, while the low-ionisation and atomic lines exhibit two peaks at farther distances. Moreover, H$_2$ lines detected with \textit{JWST} seem to peak closer to the low-ionisation and atomic lines \citep[see][]{aleman2011}. It is important to note that due to current differences in spatial resolution between MUSE and \textit{JWST}, it is difficult to assess whether [O~{\sc i}] $\lambda$6300 and H$_2$ emission lines are truly co-spatial, which would be necessary to confirm the linear relationship between the two lines presented by \cite{Reay1988}. The emission from polycyclic aromatic hydrocarbons (PAHs) also seems to be co-spatial with H$_2$ emission (Table~\ref{JWST_MUSE_e}). A similar correlation has also been presented for the Ring nebula \citep{Cox2016}. Figure~\ref{rgb_radial} shows an RGB image composed of  MUSE [O~{\sc i}] $\lambda$6300 emission as red, NIRCam@\textit{JWST} H$_2$ 2.12 $\upmu$m emission as green and MIRI@\textit{JWST} PAHs 11.3 $\upmu$m emission as blue. The image reveals that the three emissions appear to emanate from the same region, at the nebula's ring.

The region where low-ionisation ions and molecular emissions originate appears as a ring structure in two-dimensional images (Fig.~\ref{ngc3132_jwst_rgb}), similar to \citet{manchado2015}. However, the 3D model enables us to explore the three-dimensional space, where this region resembles more of a torus or a tube with a density gradient \citep[see fig.~4 in][]{hektor2025}. It is important to note that the model does not account for molecular emissions. The idea that molecules could originate from a torus in bipolar nebulae has been proposed and tested previously \citep{kastner1996, moraga2023}. In the case of NGC~3132, we expect that H$_2$ emanates from a similar torus as the low-ionisation lines, as shown by \citet{kastner2024}.

\subsection{Molecular hydrogen emission}

\begin{figure}     
    \centering
    \includegraphics[width=0.45\textwidth]{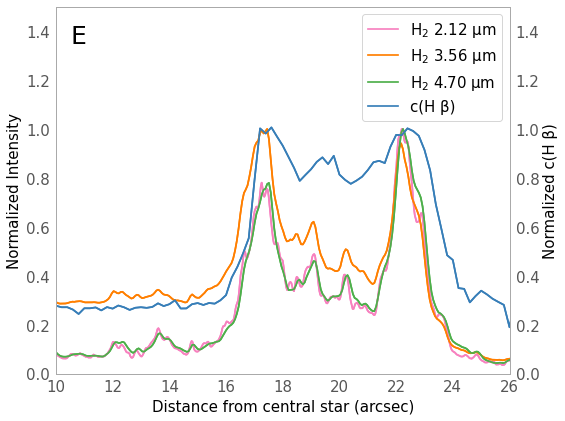}
    \caption{Radial profiles of H$_{2}$ lines and c(H~$\upbeta$) east from the CSPN.}
    \label{chb_H2}
\end{figure}

The spatial distribution of the extinction coefficient c(H~$\upbeta$), derived from the Balmer lines in MUSE data (H~$\upalpha$ and H~$\upbeta$), and H$_{2}$ lines from NIRCam@\textit{JWST}, was also explored. Remarkably, c(H~$\upbeta$) is found to reach its highest values at the same positions where H$_{2}$ lines peak, at distances $\sim$17\arcsec~and $\sim$22\arcsec~from the central star (Table~\ref{table:10}). Figure~\ref{chb_H2} presents the radial profiles of c(H~$\upbeta$) and H$_2$ emissions, normalized to one, along the eastern part of the nebula's ring. A similar analysis was also conducted for the north, west, and south directions (see Appendix~\ref{h2_chb_app}). A spatial offset of $\sim$2\arcsec~along the western side of the nebula was observed, which is likely attributed to projection effects. \cite{akras2022may} also reported an increase of c(H~$\upbeta$) at the pair of LISs \citep[low ionisation structures,][]{denise2001} in NGC~7009, where H$_2$ has also been recently detected \citep{Akrasgreen}.

\begin{table}
\centering
\caption{Radial profiles' peaks of c(H~$\upbeta$) and H$_2$ emission lines east from the CSPN.}
\resizebox{0.35\textwidth}{!}{%
\begin{tabular}{|c|c|c|} 
 \hline
 Emission line & 1$^{\rm st}$ peak & 2$^{\rm nd}$ peak \\ [0.5ex] 
 
  & (\arcsec) & (\arcsec) \\
 \hline
 c(H~$\upbeta$) & 17.2 $\pm$ 0.2 & 22.4 $\pm$ 0.2 \\
 
 H$_{2}$ 2.12 $\upmu$m & 17.26 $\pm$ 0.03 & 22.23 $\pm$ 0.03 \\
 
 H$_{2}$ 3.56 $\upmu$m & 17.45 $\pm$ 0.06 & 22.18 $\pm$ 0.06 \\
 
 H$_{2}$ 4.71 $\upmu$m & 17.51 $\pm$ 0.03 & 22.24 $\pm$ 0.03 \\
 \hline
\end{tabular}}
\label{table:10}
\end{table}

Small variations between different H$_2$ emission lines are expected due to differences in the energy required for their transitions. For instance, the H$_2$ 2.12 $\upmu$m line corresponds to 1-0 S(1) ro-vibrational transition, which requires slightly less energy than the pure rotational 0-0 S(9), 0-0 S(13) and 0-0 S(15) transitions covered by F470N and F356W filters, respectively. In contrast, the F770W filter encompasses the 0-0 S(4) and 0-0 S(5) transitions, which arise from lower energy levels compared to 1-0 S(1).

\textit{Spitzer} IRAC bands encompass H$_2$ lines from the ground rotational state 0-0 and forbidden emission lines. Channel 2 ([4.5]) contains Br~$\upalpha$ and various 0–0 H$_{2}$ emissions in the range 4.649-4.694~$\upmu$m, such as S(9) 4.69 $\upmu$m \citep{ngc3132_irac,phillips2010}. Channel 3 ([5.8]) includes S(7) 5.5 $\upmu$m and S(6) 6.1 $\upmu$m, while S(4) 8.0 $\upmu$m, S(5) 6.9 $\upmu$m and [Ar~{\sc iii}] 8.991 $\upmu$m lines are covered by channel 4 ([8.0]) \citep{Akrasgreen}. Thus, the combined information of \textit{JWST} and \textit{Spitzer}, can shed light on the link between the IRAC [8.0]/[4.5] and [5.8]/[4.5] ratios with molecular hydrogen emission. \citet{phillips2010} investigated specific ratios such as [8.0]/[4.5] and [5.8]/[4.5]. They noticed that these ratios increase at greater distances from the central stars, with [8.0]/[4.5] showing dips in low-ionisation regions outside the nebular core. \cite{Akrasgreen} also showed that the pairs of LISs in NGC~7009, where H$_2$ emission has been detected, are also characterized by strong emission in the [4.5] band, as well as low [8.0]/[4.5] and [5.8]/[4.5] ratios.

\begin{figure}   
    \centering{
    \includegraphics[width=0.45\textwidth]{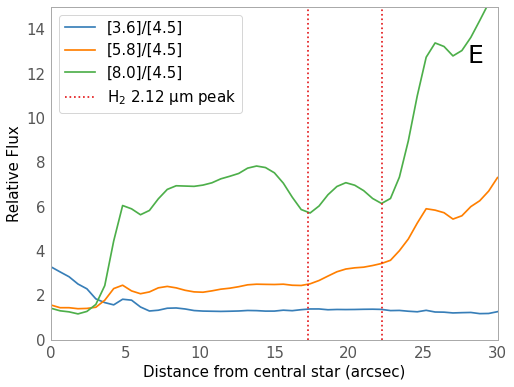}}
    \caption{Radial profile for \textit{Spitzer}'s channels ratios east from the CSPN. Red lines correspond to the local maxima of H$_2$ emission at 2.12 $\upmu$m from NIRCam@\textit{JWST}.}
    \label{spitzer_ratios}
\end{figure}

Figure~\ref{spitzer_ratios} displays the radial profiles of the IRAC [3.6]/[4.5], [4.8]/[4.5], and [8.0]/[4.5] ratios from \textit{Spitzer}\footnote{Considering as distance D=754 pc \citep{BaylerJones2021} and an expansion velocity 25~km~s$^{-1}$, the angular expansion between the \textit{Spitzer} and NIRCam observations was determined to be 0.094\arcsec~or $\sim$3.4$\times10^{-4}$ pc.} and H$_2$ 2.12 $\upmu$m line\footnote{Only, the H$_2$ 2.12 $\upmu$m emission is displayed on the diagrams, since the rest of H$_2$ are almost co-spatial. The exact distances where all the NIRCam H$_2$ emissions peak are presented in Tables~\ref{table:10}, \ref{table_h2_chb1}, \ref{table_h2_chb2}, and \ref{table_h2_chb3}} from \textit{JWST}. The comparison between NIRCam@\textit{JWST} and IRAC@\textit{Spitzer} radial profiles provides the first direct verification of the link between the H$_2$ emission lines and the [8.0]/[4.5] IRAC ratio, suggested by several authors \citep{tappe2012,quino2011}. 

The previously mentioned H$_2$ torus (see Sec.~\ref{ionic_strat}) could be traced by the connection found between the extinction coefficient c(H~$\upbeta$) and the [8.0]/[4.5] IRAC@\textit{Spitzer} ratio with H$_2$. This c(H~$\upbeta$)-H$_2$ link corroborates that dust attenuates ultraviolet radiation and protects H$_2$ from dissociation. Additionally, dust grains can also act as catalysts for the formation of H$_2$ \citep{cazaux2016,wakelam2017}.

\begin{figure*}     
    \centering{
    \includegraphics[width=1\textwidth]{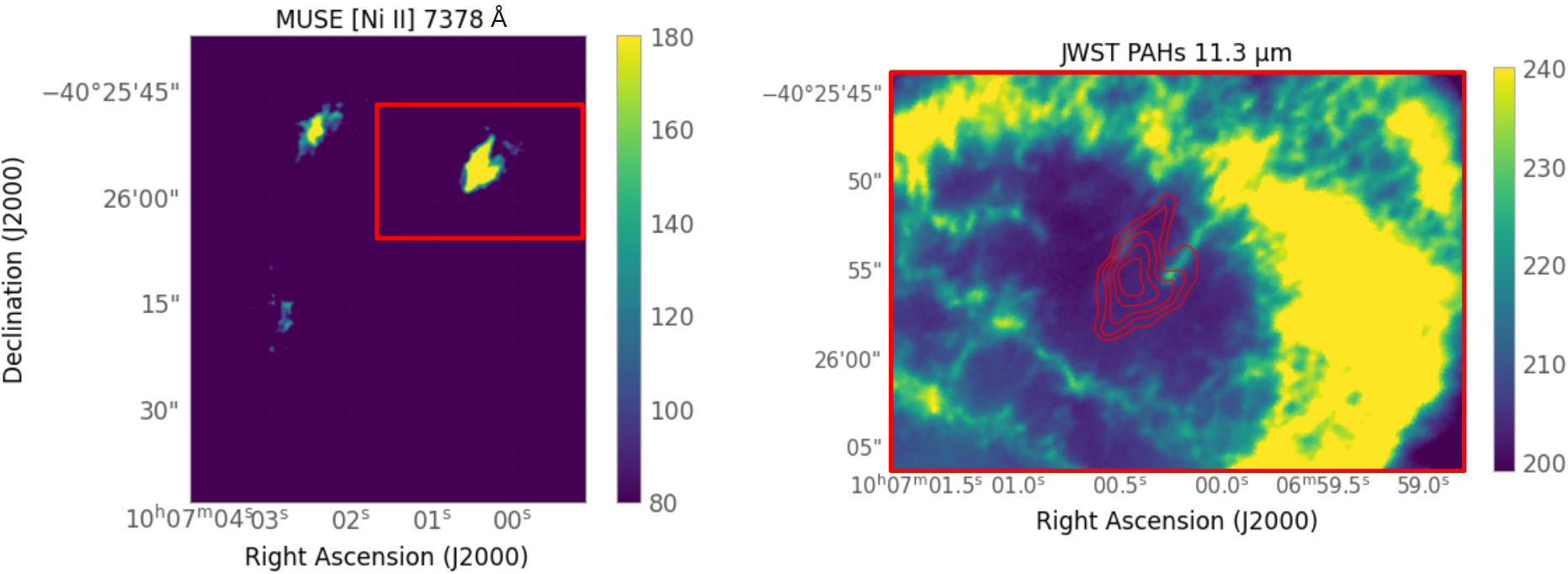}}\caption{Left panel: NGC~3132 MUSE emission line map of [Ni~{\sc ii}] 7378~\AA. Right panel: NGC~3132 MIRI@\textit{JWST} PAHs 11.3 $\upmu$m infrared emission map with red contours of [Ni~{\sc ii}] 7378~\AA~emission displayed on top of it. The red box in the [Ni~{\sc ii}] 7378~\AA~map highlights the zoomed in region shown in MIRI@\textit{JWST} PAHs 11.3 $\upmu$m emission map. colour bars correspond to the energy flux units ($\times$10$^{-20}$ erg cm$^{-2}$ s$^{-1}$).}
    \label{PAH_Ni}
\end{figure*}

\section{New discovered nickel/iron-rich clumps}
\label{nife}
By scrutinizing the MUSE datacube for weaker emission lines, we achieved the first detection of the [Ni~{\sc ii}]~$\lambda$7378, [Fe~{\sc ii}]~$\lambda$8617, and [Fe~{\sc iii}]~$\lambda$5270 lines in NGC~3132. The aforementioned emission line maps were extracted from MUSE datacube by fitting a Gaussian profile and subtracting the continuum. As a reference, the detection of the weakest emission line (i.e., [Ni~{\sc ii}]~$\lambda$7378) was achieved with a signal-to-noise ratio of four. In addition, [Ni~{\sc ii}] $\lambda$7378 relative flux to H~$\upbeta$ = 100 is approximately 0.5.

The left panel of Fig.~\ref{PAH_Ni} presents the first spatially resolved map of the [Ni~{\sc ii}]~$\lambda$7378 emission line, while the right panel shows the contours of [Ni~{\sc ii}]~$\lambda$7378 line overlaid on the PAH 11.3 $\upmu$m emission map from MIRI@\textit{JWST}. Notably, several smaller H$_2$ cometary knots seem to emanate from the same region (Fig.~\ref{jwst_h2}).

It should be noted that the newly discovered Ni(Fe)-rich clumps are likely associated with a north-western cometary knot detected in the F1130W MIRI image and H$_2$ cometary knots present in F212N NIRCam image. The clump displays a V-shaped structure, indicative of an interaction with the knot. Ni(Fe)-rich clump appears to have been ejected at a later time compared to PAH-H$_2$ cometary knot, with the latter being more dense as it deforms the Ni(Fe)-rich clump. In addition, the PAH-H$_2$ cometary knot share similarities with cometary knots in Helix nebula \citep{meaburn1998, matsuura2009}, whose formation has been attributed to various mechanisms, such as in-situ formation \citep{segura2006} or origin during the AGB phase \citep{sokerr1998}. The enhanced Ni (and Fe) emission in these clumpy structures suggests dust destruction by shock waves. Further investigation is required to understand these previously unknown regions and the origin of their emission. A comprehensive study has been developed, encompassing a broader sample of PNe (Bouvis et al. 2025, submitted).

\section{Conclusions}

\label{summary}
A detailed analysis of a sophisticated 3D photoionisation model of the planetary nebula NGC~3132 was conducted utilizing {\sc satellite} code, along with MUSE, \textit{JWST}, and \textit{Spitzer} observations. The key findings of this study are summarized as follows:

\begin{enumerate}
    \item Physical parameters derived from emission line fluxes of the most advanced 3D photoionization model available in literature, such as $T_{\rm e}$, $n_{\rm e}$, and their spatial distribution are consistent with MUSE data. However, some ionic and total elemental abundances deviate from the observational results.
    \item Pseudo-slits that do not encompass all ionization zones, as well as individual spaxels, tend to produce inaccurate (higher) chemical abundances.
    \item Diagnostic diagrams derived from observations and the model show good agreement, confirming that NGC~3132 is a high-ionisation nebula. Its outer regions are dominated by low-ionisation gas, and no shock excitation is required to reproduce the observations. 
    \item Employing the 3D photoionization model and simulated observations, we find that the empirical method for estimating elemental abundances fails to recover the original (input) abundances. Specifically for oxygen, a 35$\%$ overestimation of the total abundance is measured. This discrepancy is attributed mainly to projection effects on the fluxes and $T_{\rm e}$, with the ICF formulae playing a minor role.
    \item The ionisation structure of the nebula was examined through the radial profiles of optical emission lines from MUSE and infrared emission lines from \textit{JWST}. Low-ionisation and atomic lines appear to be co-spatial with H$_2$ emission. Additionally, our observational analysis suggests a link between [O~{\sc i}]~$\lambda$6300 and H$_2$ 2.12 $\upmu$m lines and PAHs features.
    \item  We propose that H$_2$ originates from a dusty torus surrounding the PN, as previously suggested for bipolar PNe. The existence of this torus is further supported by the correlation between H$_2$ emission, dust traced by c(H~$\upbeta$), and the dips in the IRAC [8.0]/[4.5] ratio. This further supports the link between dust and H$_2$ emission, with the former likely acting as a catalyst for H$_2$ formation. 
    \item Three new Ni(Fe)-rich clumps are identified in the MUSE data of NGC~3132 for the first time. The northwest clump appears to interact with a PAH-H$_2$ cometary knot visible in \textit{JWST} images, giving it a V-shape. The detection of Ni (and Fe) emission suggest the existence of shock waves that destroy dust grains, releasing these elements into the gas phase.  
\end{enumerate}

\section*{Acknowledgements}

We sincerely thank the referee, Christophe Morisset, for his thorough review and valuable suggestions, which improved the clarity and quality of this paper. The research project is implemented in the framework of H.F.R.I. call “Basic research Financing (Horizontal support of all Sciences)" under the National Recovery and Resilience Plan “Greece 2.0" funded by the European Union – NextGenerationEU (H.F.R.I. Project Number: 15665). JGR acknowledges financial support from the Agencia Estatal de Investigaci\'{o}n of the Ministerio de Ciencia e Innovaci\'{o}n (AEI- MCINN) under Severo Ochoa Centres of Excellence Programme 2020-2023 (CEX2019-000920-S), and from grant PID-2022136653NA-I00 (DOI:10.13039/501100011033) funded by the Ministerio de Ciencia, Innovaci\'{o}n y Universidades (MCIU/AEI) and by ERDF "A way of making Europe" of the European Union. DRG acknowledges FAPERJ (E-26/211.527/2023) and CNPq (315307/2023-4). 

This work has made use of the computing facilities available at the Laboratory of Computational Astrophysics of the Universidade Federal de Itajub\'{a} (LAC-UNIFEI). The LAC-UNIFEI is maintained with grants from CAPES, CNPq and FAPEMIG.

\section*{Data Availability}

The data and results of this article will be shared on reasonable request to the corresponding author. The {\sc satellite} code, along with its documentation and examples, is available from the GitHub repository [\url{https://github.com/StavrosAkras/SATELLITE.git}].



\bibliographystyle{mnras}
\bibliography{biblio} 



\clearpage
\appendix

\section{Complementary figures}
\subsection{Specific slit}
\label{app_spec_slit}

\begin{figure}    
    \centering
    \includegraphics[width=0.43\textwidth]{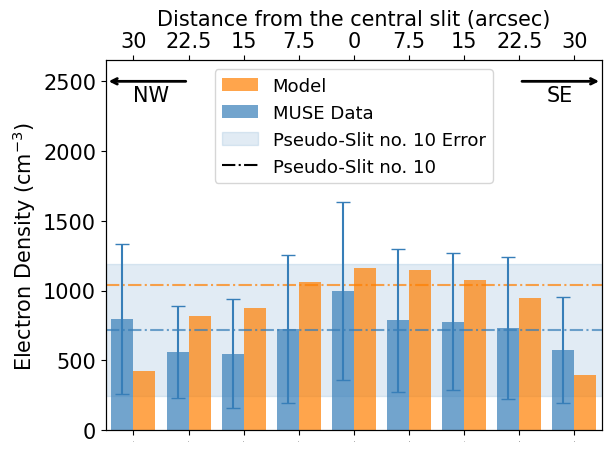}
    \includegraphics[width=0.43\textwidth]{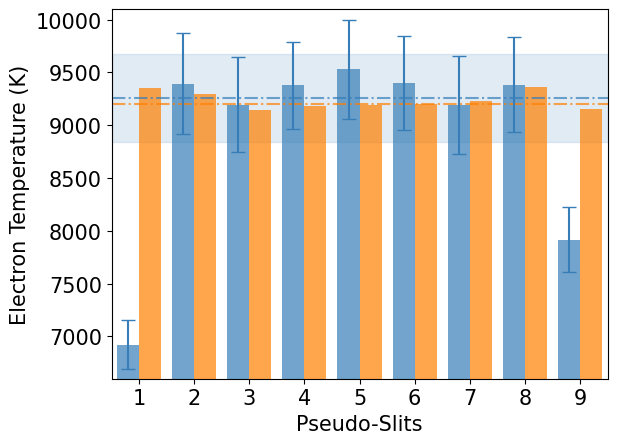}
    \caption{Top panel: $n_{\rm e}$ [Cl~{\sc iii}] for each pseudo-slit. Bottom panel: $T_{\rm e}$ [S~{\sc iii}] for each pseudo-slit. Blue colour denotes MUSE data, while orange model values.}
    \label{ne_cl3}
\end{figure}

\begin{figure}    
    \centering
    \includegraphics[width=0.43\textwidth]{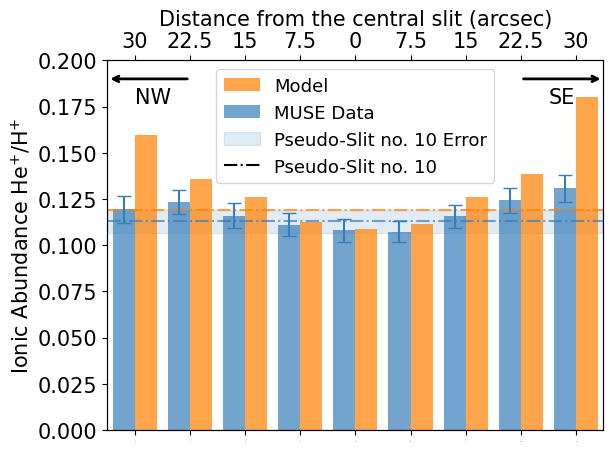}
    \includegraphics[width=0.43\textwidth]{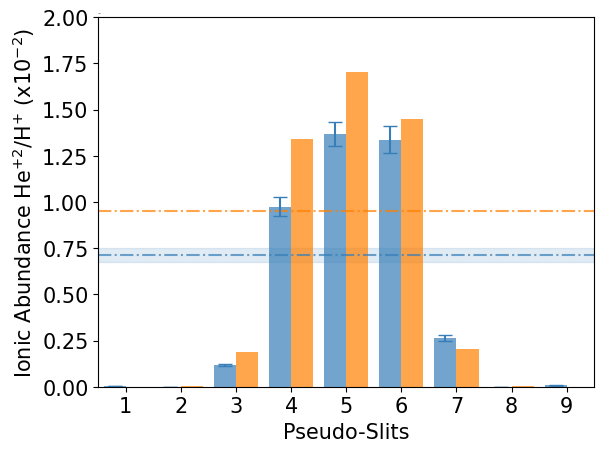}
    \caption{Ionic abundance for He$^+$ and He$^{+2}$. MUSE data are denoted with blue colour, while 3D model values with orange.}
    \label{He_ion_abund}
\end{figure}

\begin{figure}    
    \centering
    \includegraphics[width=0.43\textwidth]{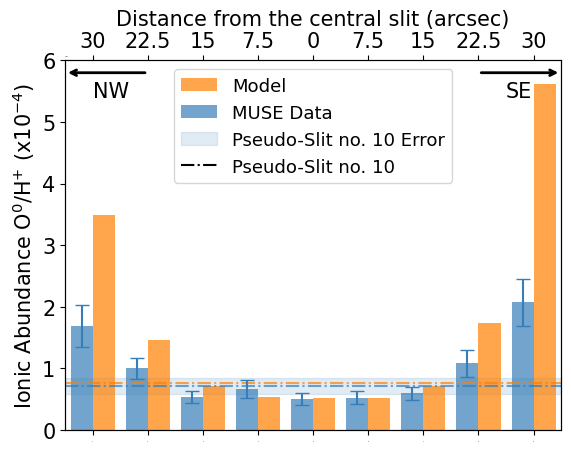}
    \includegraphics[width=0.43\textwidth]{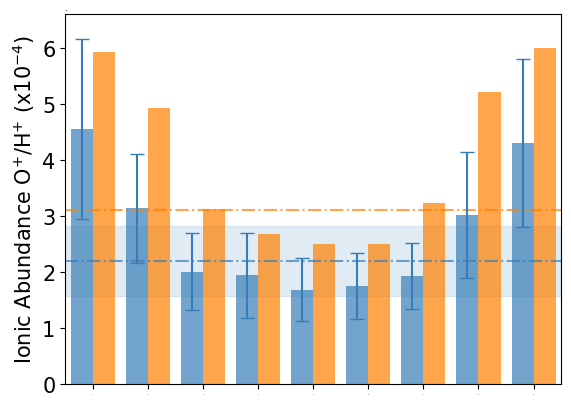}
    \includegraphics[width=0.43\textwidth]{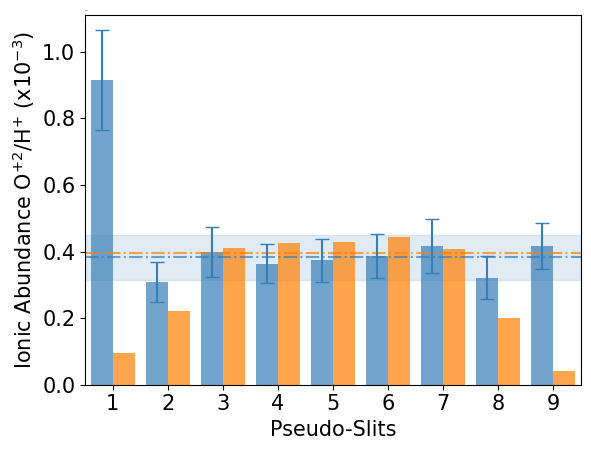}
    \caption{Atomic abundance of O$^0$ and ionic abundances of O$^+$ and O$^{+2}$. MUSE data are denoted with blue colour, while 3D model values with orange.}
    \label{O_ion_abund}
\end{figure}

\begin{figure}    
    \centering
    \includegraphics[width=0.43\textwidth]{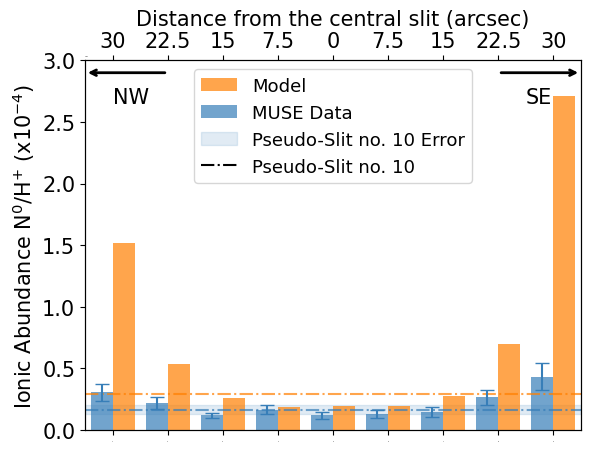}
    \includegraphics[width=0.43\textwidth]{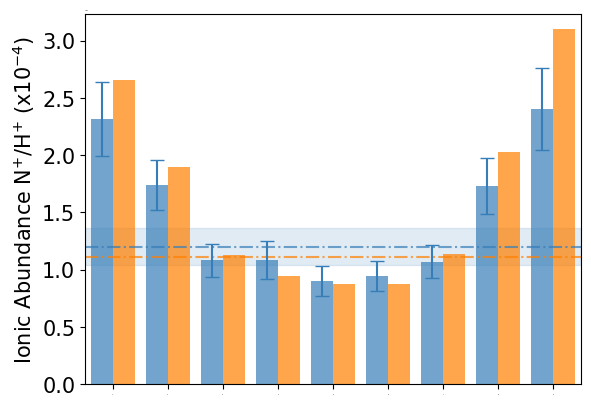}
    \includegraphics[width=0.43\textwidth]{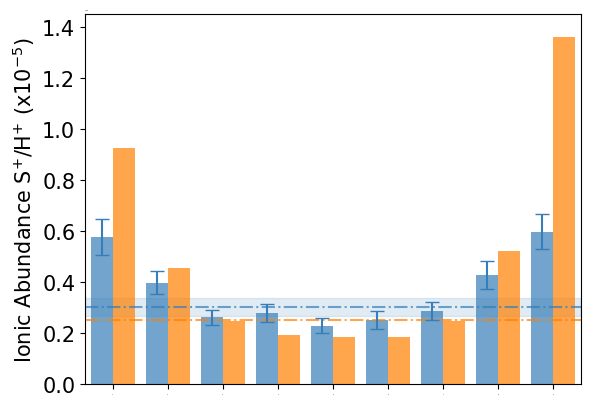}
    \includegraphics[width=0.43\textwidth]{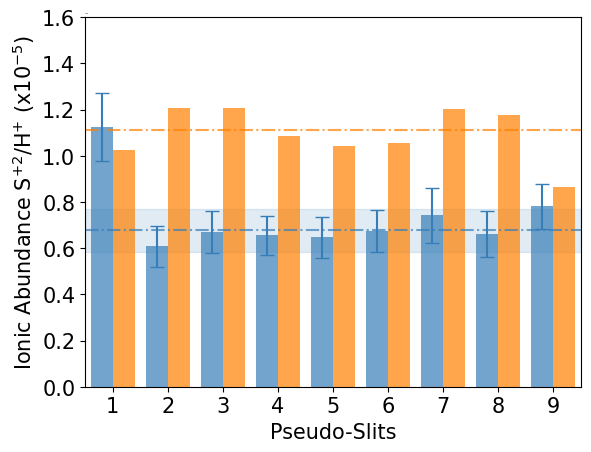}
    \caption{Atomic abundance of N$^0$ and ionic abundances of N$^+$, S$^+$ and S$^{+2}$. MUSE data are denoted with blue colour, while 3D model values with orange.}
    \label{N_S_ion_abund}
\end{figure}

\begin{figure}    
    \centering
    \includegraphics[width=0.43\textwidth]{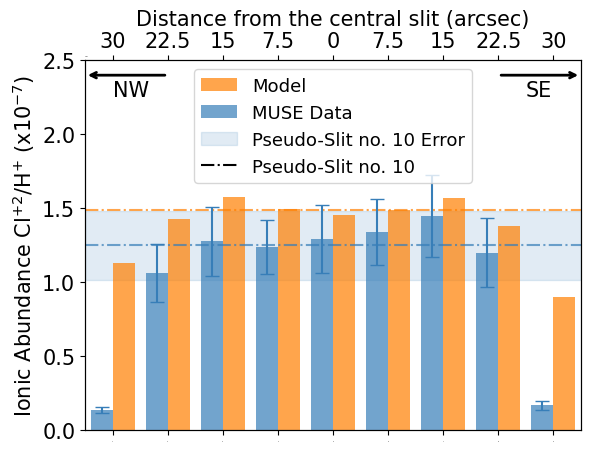}
    \includegraphics[width=0.43\textwidth]{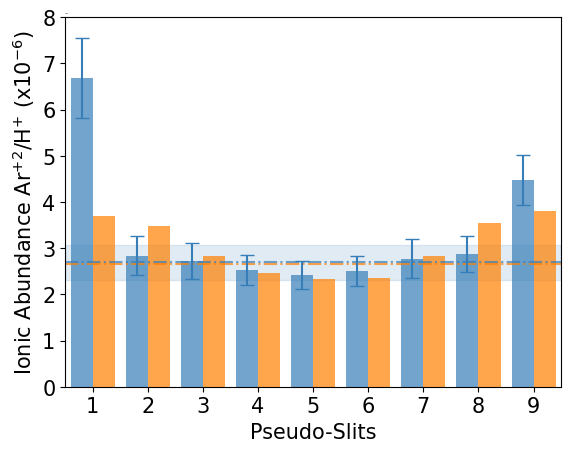}
    \caption{Ionic abundance for Cl$^{+2}$ and Ar$^{+2}$. MUSE data are denoted with blue colour, while 3D model values with orange.}
    \label{Cl_Ar_ion_abund}
\end{figure}

\begin{figure}    
    \centering
    \includegraphics[width=0.43\textwidth]{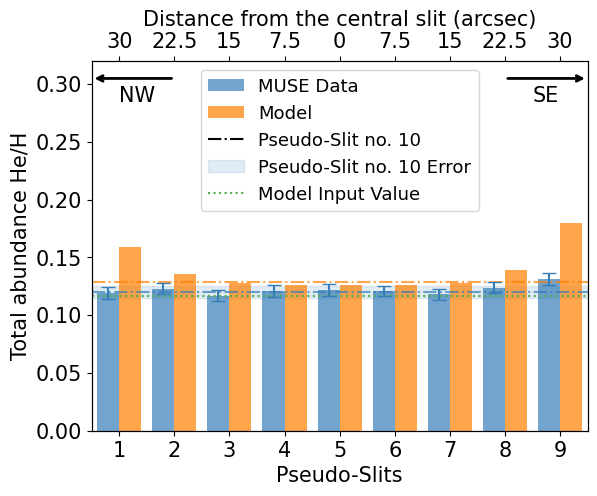}
    \caption{Total He abundance as estimated by MUSE data (blue) and the 3D model (orange). The dotted green line corresponds to the abundance that was used as input for the model.}
    \label{he_abund}
\end{figure}

\begin{figure}    
    \centering
    \includegraphics[width=0.43\textwidth]{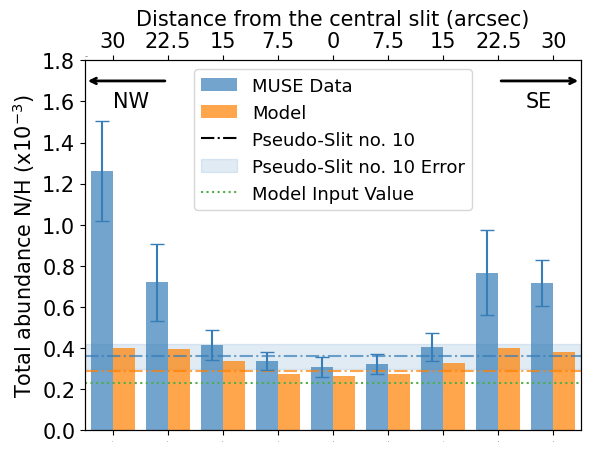}
    \includegraphics[width=0.43\textwidth]{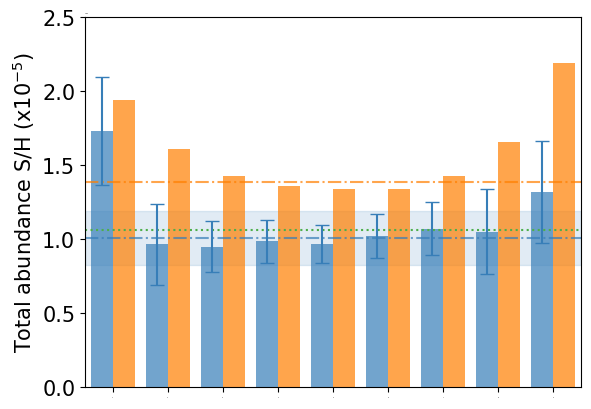}
    \includegraphics[width=0.43\textwidth]{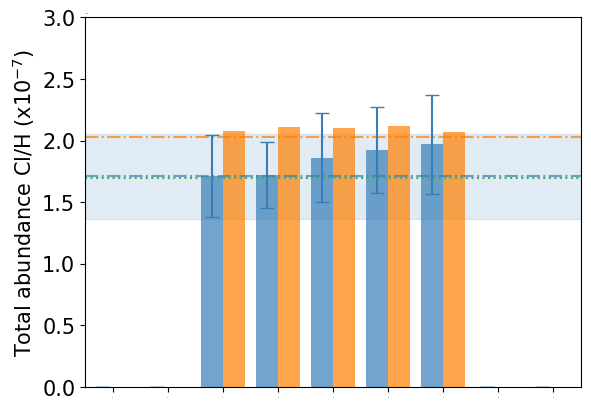}
    \includegraphics[width=0.43\textwidth]{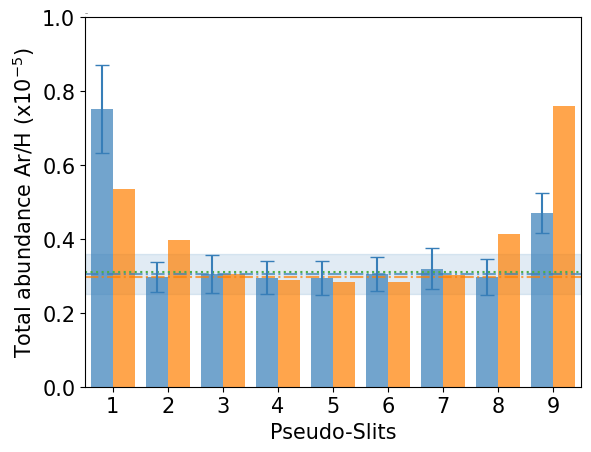}
    \caption{Total N, S, Cl and Ar abundances estimated with ICF from DIMS14. MUSE data are denoted with blue colour, while 3D model values with orange. The dotted green line corresponds to the abundance that was used as input for the model.}
    \label{rest_abund}
\end{figure}

\begin{figure}    
    \centering
    \includegraphics[width=0.43\textwidth]{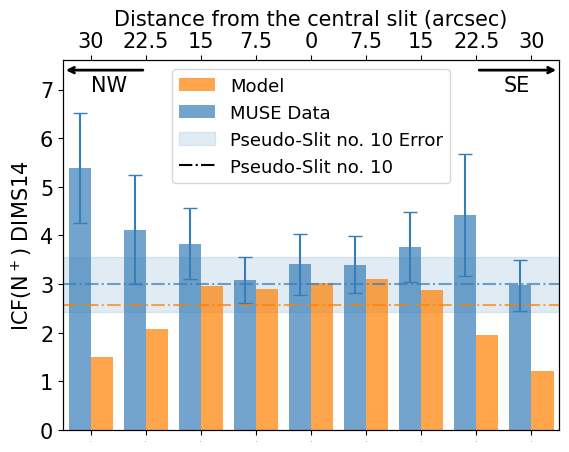}
    \includegraphics[width=0.43\textwidth]{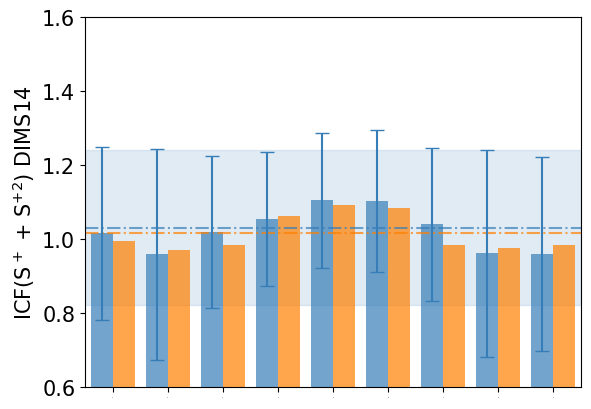}
    \includegraphics[width=0.43\textwidth]{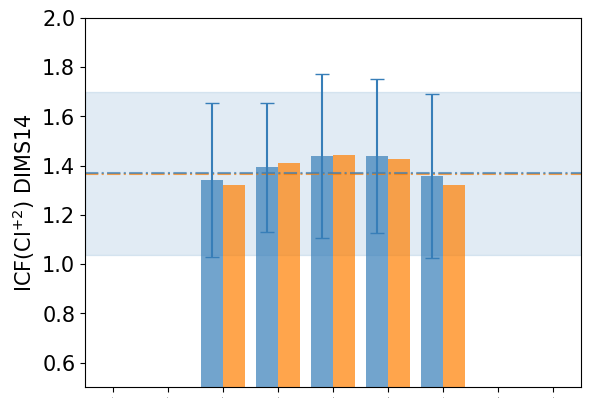}
    \includegraphics[width=0.43\textwidth]{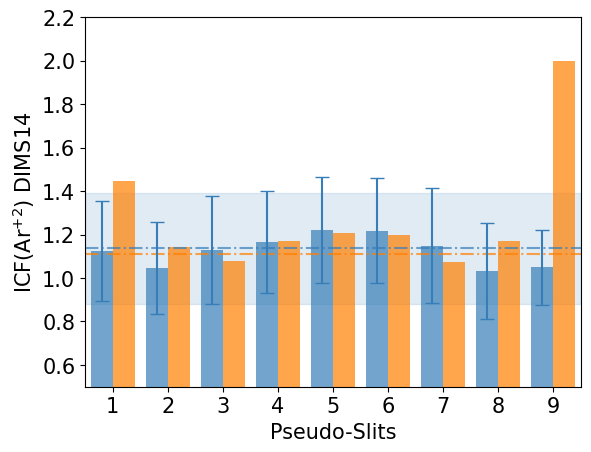}
    \caption{ICF estimates from DIMS14 for N, S, Cl and Ar. MUSE data are denoted with blue colour, while 3D model values with orange.}
    \label{rest_ICFs}
\end{figure}

\subsection{ICF maps}
\begin{figure}    
    \centering
    \includegraphics[width=0.43\textwidth]{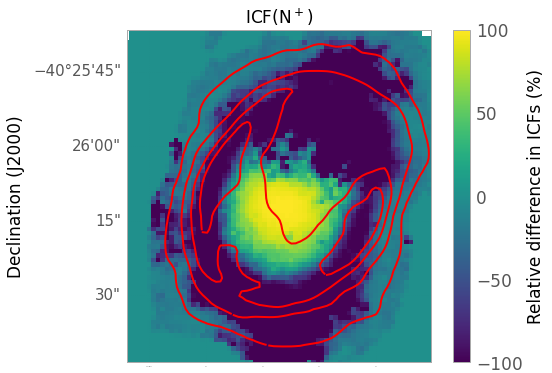}
    \includegraphics[width=0.43\textwidth]{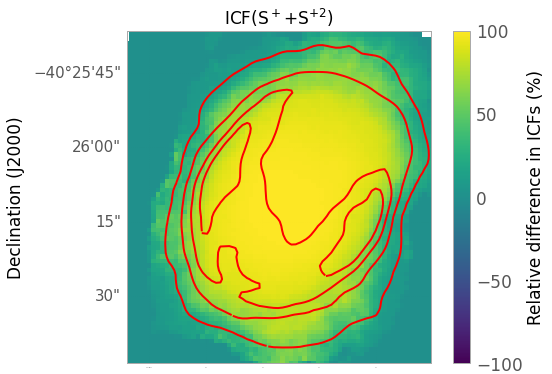}
    \includegraphics[width=0.43\textwidth]{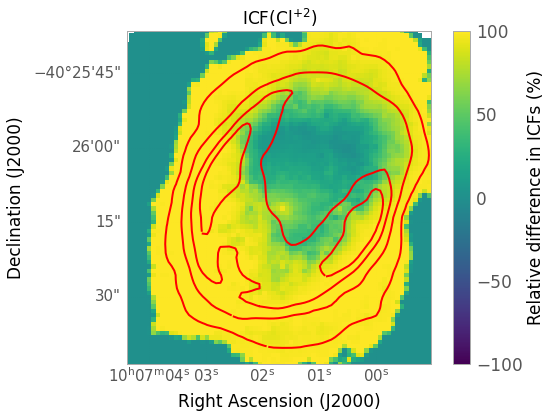}
    \caption{2D maps of the relative difference in ICFs for N, S and Cl derived from the 3D model and the one using the formulae of DIMS14. The red contours indicate the surface brightness of the H $\upbeta$ emission line.}
    \label{rest_2d_ICFs}
\end{figure}

\clearpage

\subsection{H$_2$ and c(H~$\upbeta$) correlation}
\label{h2_chb_app}

\begin{figure}    
    \centering
    \includegraphics[width=0.48\textwidth]{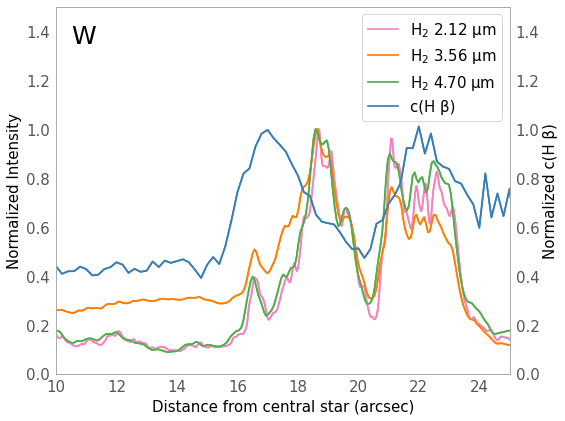}
    \includegraphics[width=0.48\textwidth]{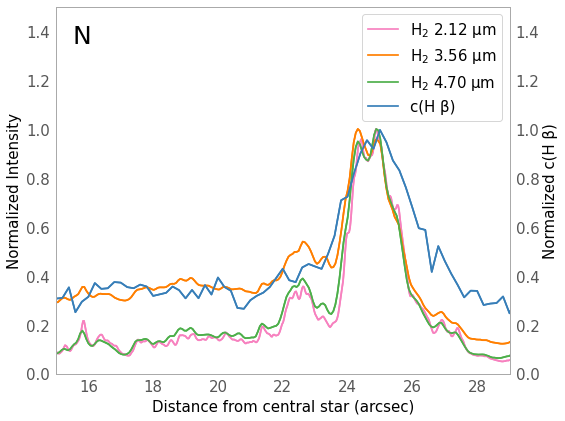}
    \includegraphics[width=0.48\textwidth]{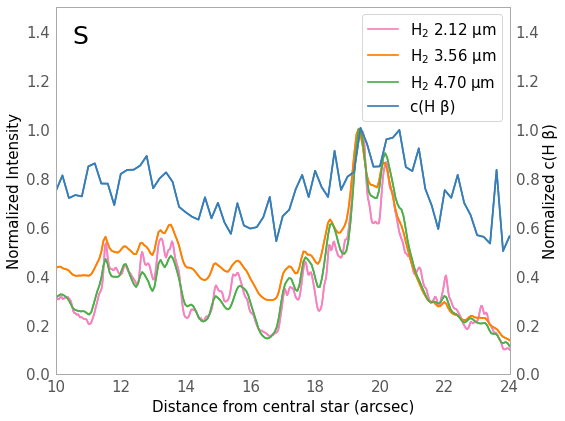}
    \caption{Radial profiles of H$_{2}$ lines and extinction coefficient west, north and south from the CSPN, respectively.}
    \label{H2_CHB}
\end{figure}

\subsection{IRAC colours and H$_2$}
\label{irac_h2_app}

\begin{figure}    
    \centering{
    \includegraphics[width=0.48\textwidth]{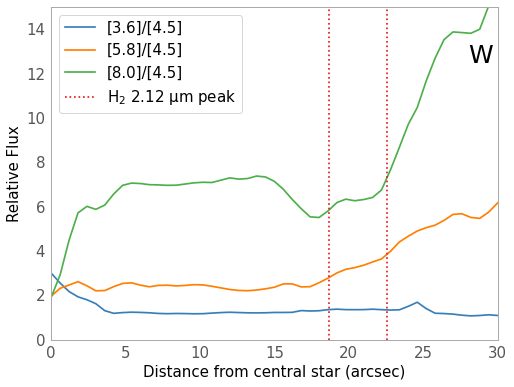}
    \includegraphics[width=0.48\textwidth]{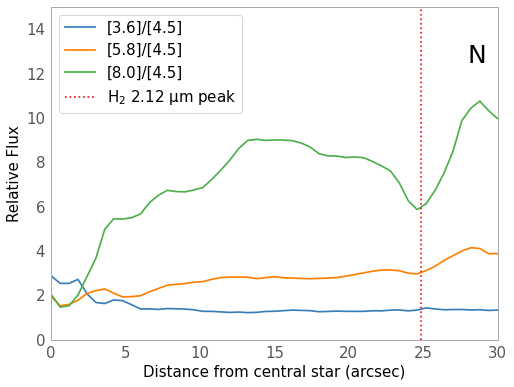}
    \includegraphics[width=0.48\textwidth]{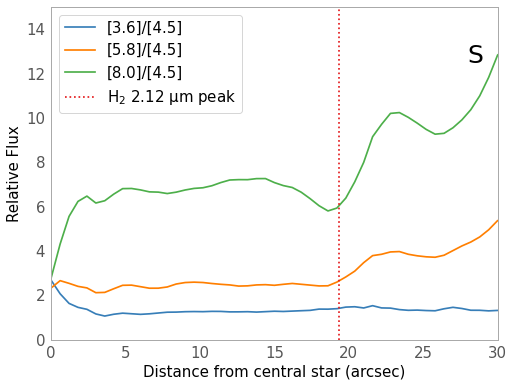}}
    \caption{Radial profile for \textit{Spitzer}'s channels ratios west, north and south from the CSPN, respectively. Red lines correspond to the local maxima of H$_2$ emission at 2.12 $\upmu$m from NIRCam@\textit{JWST}.}
    \label{IRAC_H2}
\end{figure}

\clearpage

\subsection{SATELLITE radial analysis}
\label{sat_rad_app}
    
\begin{figure}
    \centering
    \includegraphics[width=0.48\textwidth]{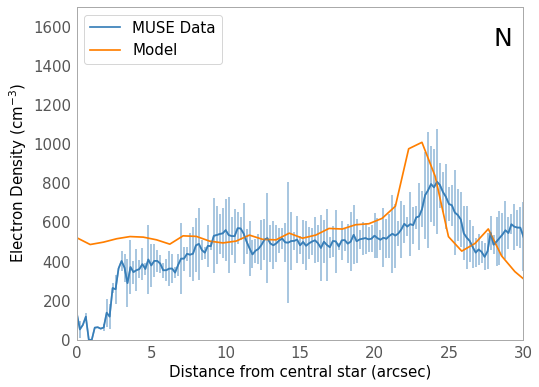}
    \includegraphics[width=0.48\textwidth]{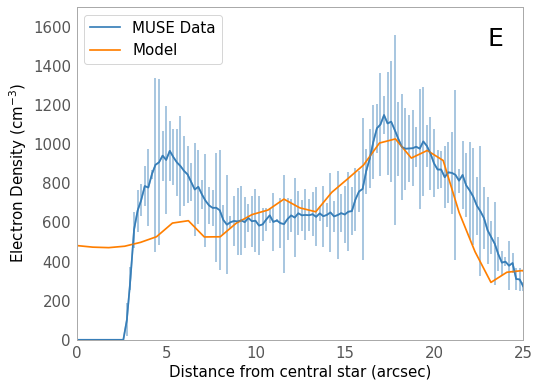}
    \includegraphics[width=0.48\textwidth]{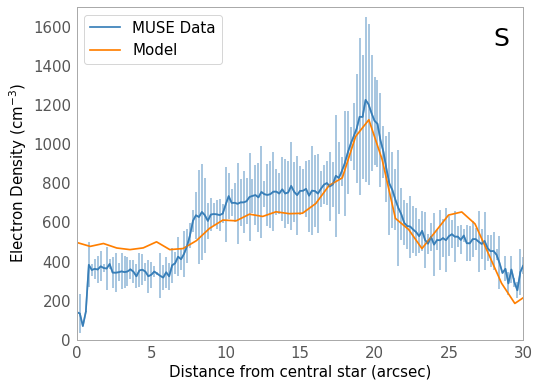}
    \caption{Radial profile of $n_{\rm e}$ [S~{\sc ii}] north, east and south from the CSPN, respectively, as estimated from the MUSE data (blue) and the model (orange).} 
    \label{ne_rad_north}
\end{figure}

\begin{figure}
    \centering
    \includegraphics[width=0.48\textwidth]{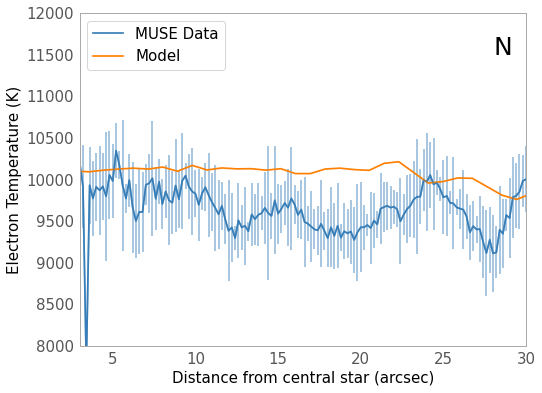}
    \includegraphics[width=0.48\textwidth]{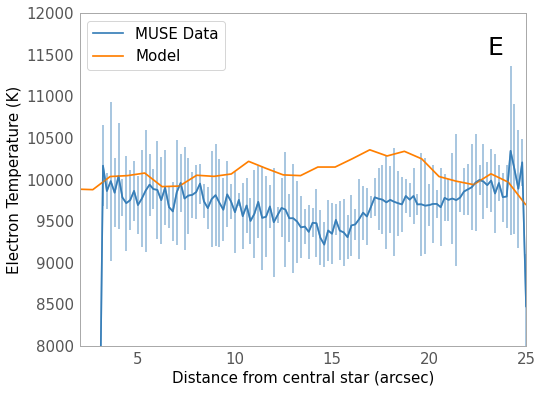}
    \includegraphics[width=0.48\textwidth]{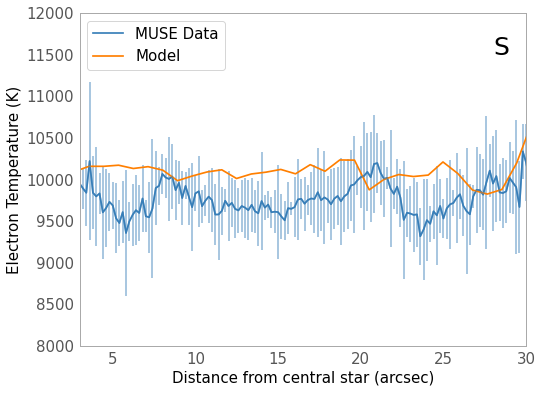}
    \caption{Radial profile of $T_{\rm e}$ [N~{\sc ii}] north, east and south from the CSPN, respectively, as estimated from MUSE data (blue) and the model (orange).} 
    \label{te_rad_north}
\end{figure}

\begin{figure}
    \centering
    \includegraphics[width=0.5\textwidth]{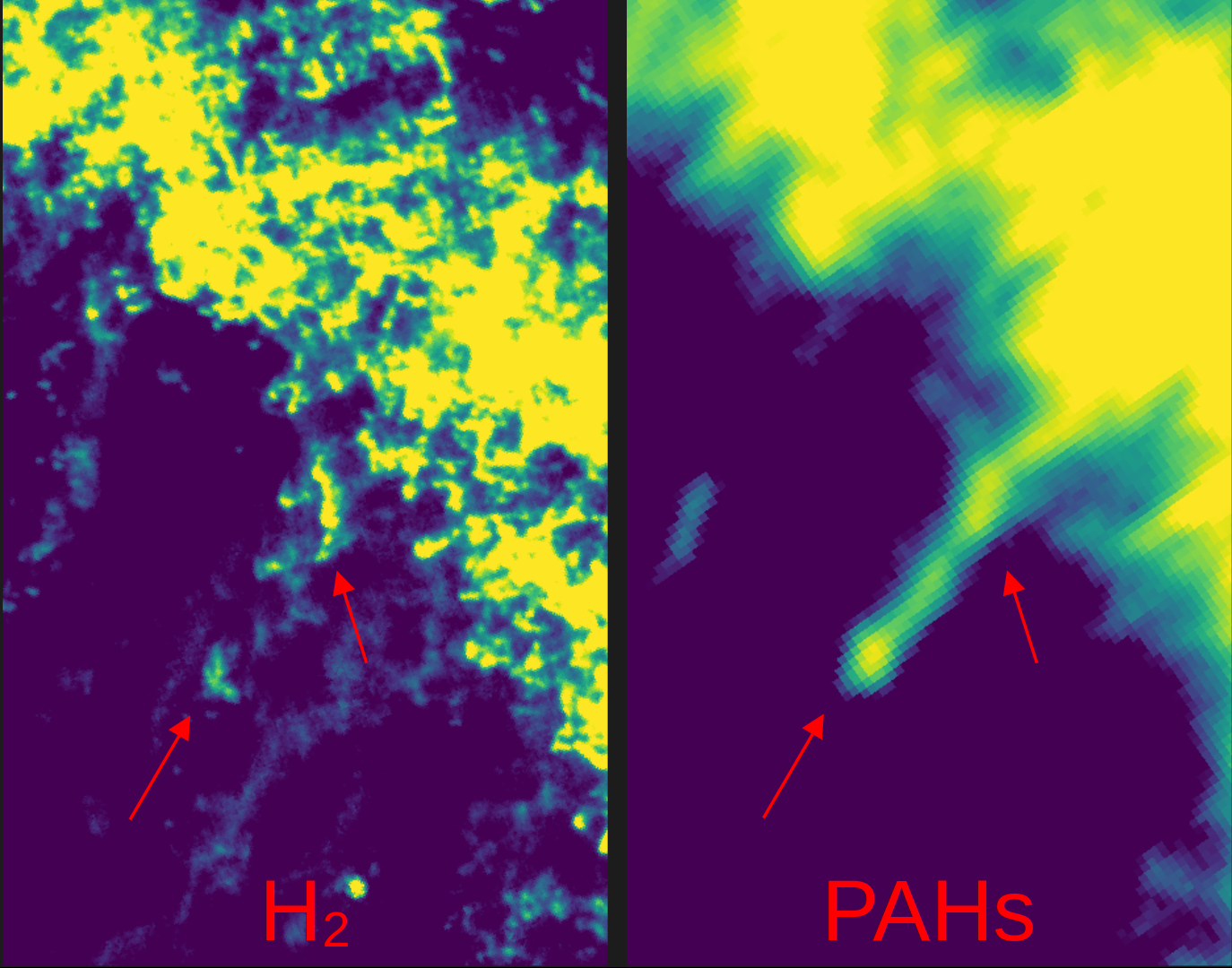}
    \caption{Left panel: NIRCam@\textit{JWST} H$_2$ 2.12 $\upmu$m emission observed with NIRCam@\textit{JWST}, highlighting the correlation between H$_2$ and PAHs in two cometary knots (red arrows). Right Panel: PAHs emission at 11.3 $\upmu$m as captured by MIRI@\textit{JWST}.} 
    \label{jwst_h2}
\end{figure}

\section{Complementary tables}

\label{jwst_muse_app}

\begin{table}
\centering
\caption{Radial profiles' peaks of MUSE and \textit{JWST} emissions west from the CSPN.}
\resizebox{0.4\textwidth}{!}{%
\begin{tabular}{|c|c|c|} 
 \hline
 Emission line & 1$^{\rm st}$ peak & 2$^{\rm nd}$ peak \\ [0.2ex] 
 
  & (\arcsec) & (\arcsec) \\
 \hline
 [O~{\sc iii}] 5007~\AA & 17.6 $\pm$ 0.2 & - \\ 
 
 [S~{\sc iii}] 9069~\AA & 17.6 $\pm$ 0.2 & - \\
 
 [S~{\sc iii}] 9069~\AA~(\textit{JWST}) & 17.84 $\pm$ 0.03 & 21.0 $\pm$ 0.03 \\
 
 Pa~$\upalpha$ & 17.92 $\pm$ 0.03 & 21.0 $\pm$ 0.03 \\
 
 [S~{\sc iii}] 18 $\upmu$m & 17.97 $\pm$ 0.11 & - \\
 
 H~$\upalpha$ 6563~\AA & 18.0 $\pm$ 0.2 & - \\ 
 
 Br~$\upalpha$ & 18.08 $\pm$ 0.06 & 21.04 $\pm$ 0.06 \\
 
 H~$\upbeta$ 4861~\AA & 18.2 $\pm$ 0.2 & - \\
 
 [N~{\sc ii}] 5755~\AA & 18.4 $\pm$ 0.2 & 20.6 $\pm$ 0.2 \\
  
 [N~{\sc ii}] 6584~\AA & 18.4 $\pm$ 0.2 & 20.6 $\pm$ 0.2 \\ 
 
 [S~{\sc ii}] 6731~\AA & 18.4 $\pm$ 0.2 & 20.8 $\pm$ 0.2 \\
 
 H$_{2}$ 4.70 $\upmu$m & 18.58 $\pm$ 0.06 & 21.04 $\pm$ 0.06 \\
 
 [O~{\sc i}] 6300~\AA & 18.6 $\pm$ 0.2 & 21.0 $\pm$ 0.2 \\
 
 H$_{2}$ 3.56 $\upmu$m & 18.65 $\pm$ 0.06 & 21.10 $\pm$ 0.06 \\
 
 H$_{2}$ 2.12 $\upmu$m & 18.68 $\pm$ 0.03 & 21.08 $\pm$ 0.03 \\
 
 [N~{\sc i}] 5199~\AA & 18.8 $\pm$ 0.2 & 21.0 $\pm$ 0.2 \\
 
 H$_{2}$ 7.7 $\upmu$m & 18.96 $\pm$ 0.11 & - \\
 
 [Ne~{\sc ii}] 12.8 $\upmu$m & 18.96 $\pm$ 0.11 & - \\
 
 PAHs 11.3 $\upmu$m & 19.0 $\pm$ 0.11 & - \\
 \hline
\end{tabular}
}
\label{JWST_MUSE_w}
\end{table}

\begin{table}
\centering
\caption{Same as Table~\ref{JWST_MUSE_w} north from the CSPN.}
\resizebox{0.4\textwidth}{!}{%
\begin{tabular}{|c|c|c|} 
 \hline
 Emission line & 1$^{\rm st}$ peak & 2$^{\rm nd}$ peak \\ [0.2ex] 
 
  & (\arcsec) & (\arcsec) \\
 \hline 
 [O~{\sc iii}] 5007~\AA & 18.4 $\pm$ 0.2 & - \\ 
 
 H~$\upbeta$ 4861~\AA & 18.4 $\pm$ 0.2 & 23.8 $\pm$ 0.2 \\
 
 [S~{\sc iii}] 9069~\AA & 18.6 $\pm$ 0.2 & 23.8 $\pm$ 0.2 \\
 
 H~$\upalpha$ 6563~\AA & 18.4 $\pm$ 0.2 & 24.0 $\pm$ 0.2 \\ 
 
 [N~{\sc ii}] 6584~\AA & - & 24.0 $\pm$ 0.2 \\ 
 
 [N~{\sc ii}] 5755~\AA & - & 24.0 $\pm$ 0.2 \\
  
 [S~{\sc iii}] 18 $\upmu$m & - & 24.07 $\pm$ 0.11 \\
 
 Br~$\upalpha$ & 18.58 $\pm$ 0.06 & 24.13 $\pm$ 0.06 \\
 
 Pa~$\upalpha$ & 18.60 $\pm$ 0.03 & 24.19 $\pm$ 0.03 \\
 
 [S~{\sc iii}] 9069~\AA~(\textit{JWST}) & 18.52 $\pm$ 0.03 & 24.22 $\pm$ 0.03 \\
 
 H$_{2}$ 4.70 $\upmu$m & - & 24.32 $\pm$ 0.06 \\
 
 H$_{2}$ 3.56 $\upmu$m & - & 24.32 $\pm$ 0.06 \\
 
 PAHs 11.3 $\upmu$m & - & 24.4 $\pm$ 0.11 \\
 
 H$_{2}$ 7.7 $\upmu$m & - & 24.4 $\pm$ 0.11 \\
 
 [Ne~{\sc ii}] 12.8 $\upmu$m & - & 24.4 $\pm$ 0.11 \\
 
 [S~{\sc ii}] 6731~\AA & - & 24.4 $\pm$ 0.2 \\
 
 H$_{2}$ 2.12 $\upmu$m & - & 24.42 $\pm$ 0.03 \\
 
 [O~{\sc i}] 6300~\AA & - & 24.6 $\pm$ 0.2 \\

 [N~{\sc i}] 5199~\AA & - & 24.6 $\pm$ 0.2 \\
 \hline
\end{tabular}
}
\label{JWST_MUSE_n}
\end{table}

\begin{table}
\centering
\caption{Same as Table~\ref{JWST_MUSE_w} south from the CSPN.}
\resizebox{0.4\textwidth}{!}{%
\begin{tabular}{|c|c|c|} 
 \hline
 Emission line & 1$^{\rm st}$ peak & 2$^{\rm nd}$ peak \\ [0.2ex] 
 
  & (\arcsec) & (\arcsec) \\
 \hline 
 [O~{\sc iii}] 5007~\AA & 18.8 $\pm$ 0.2 & - \\ 
 
 Pa~$\upalpha$ & 19.21 $\pm$ 0.03 & - \\
 
 Br~$\upalpha$ & 19.28 $\pm$ 0.06 & - \\
 
 [S~{\sc iii}] 18 $\upmu$m & 19.30 $\pm$ 0.11 & - \\
 
 [S~{\sc iii}] 9069~\AA~(\textit{JWST}) & 19.32 $\pm$ 0.03 & - \\
 
 H$_{2}$ 3.56 $\upmu$m & 19.34 $\pm$ 0.06 & 20.16 $\pm$ 0.06 \\
 
 H$_{2}$ 2.12 $\upmu$m & 19.35 $\pm$ 0.03 & 20.19 $\pm$ 0.03 \\
 
 [S~{\sc iii}] 9069~\AA & 19.4 $\pm$ 0.2 & - \\
 
 H~$\upalpha$ 6563~\AA & 19.4 $\pm$ 0.2 & - \\ 
 
 H~$\upbeta$ 4861~\AA & 19.4 $\pm$ 0.2 & - \\
 
 H$_{2}$ 4.70 $\upmu$m & 19.40 $\pm$ 0.06 & 20.16 $\pm$ 0.06 \\
 
 [Ne~{\sc ii}] 12.8 $\upmu$m & 19.52 $\pm$ 0.11 & 20.29 $\pm$ 0.11 \\
 
 H$_{2}$ 7.7 $\upmu$m & 19.52 $\pm$ 0.11 & 20.63 $\pm$ 0.11 \\
 
 [N~{\sc ii}] 6584~\AA & 19.8 $\pm$ 0.2 & - \\ 
 
 [N~{\sc ii}] 5755~\AA & 20.0 $\pm$ 0.2 & - \\
  
 [S~{\sc ii}] 6731~\AA & 20.2 $\pm$ 0.2 & - \\
 
 [O~{\sc i}] 6300~\AA & 20.4 $\pm$ 0.2 & - \\
  
 [N~{\sc i}] 5199~\AA & 20.6 $\pm$ 0.2 & - \\
 
 PAHs 11.3 $\upmu$m & 20.74 $\pm$ 0.11 & - \\
 \hline
\end{tabular}
}
\label{JWST_MUSE_s}
\end{table}

\begin{table}
\centering
\caption{Radial analysis comparison between c(H~$\upbeta$) and H$_2$ NIRCam@\textit{JWST} emissions south from the CSPN.}
\resizebox{0.35\textwidth}{!}{%
\begin{tabular}{|c|c|c|} 
 \hline
 Emission line & 1$^{\rm st}$ peak & 2$^{\rm nd}$ peak \\ [0.2ex] 
 
  & (\arcsec) & (\arcsec) \\
 \hline 
 H$_{2}$ 3.56 $\upmu$m & 19.34 $\pm$ 0.06 & 20.10 $\pm$ 0.06 \\
 
 H$_{2}$ 2.12 $\upmu$m & 19.35 $\pm$ 0.03 & 20.19 $\pm$ 0.03 \\
 
 c(H~$\upbeta$) & 19.4 $\pm$ 0.2 & 20.2 $\pm$ 0.2 \\
 
 H$_{2}$ 4.70 $\upmu$m & 19.40 $\pm$ 0.06 & 20.16 $\pm$ 0.06 \\
 \hline
\end{tabular}
}
\label{table_h2_chb1}
\end{table}

\begin{table}
\centering
\caption{Same as Table~\ref{table_h2_chb1} west from the CSPN.}
\resizebox{0.35\textwidth}{!}{%
\begin{tabular}{|c|c|c|} 
 \hline
 Emission line & 1$^{\rm st}$ peak & 2$^{\rm nd}$ peak \\ [0.2ex] 
 
  & (\arcsec) & (\arcsec) \\
 \hline 
  c(H~$\upbeta$) & 17.0 $\pm$ 0.2 & 22.4 $\pm $ 0.2 \\
 
  H$_{2}$ 4.70 $\upmu$m & 18.58 $\pm$ 0.06 & 22.49 $\pm$ 0.06 \\
 
 H$_{2}$ 3.56 $\upmu$m & 18.65 $\pm$ 0.06 & 22.55 $\pm$ 0.06 \\
 
 H$_{2}$ 2.12 $\upmu$m & 18.68 $\pm$ 0.03 & 22.60 $\pm$ 0.03 \\
 \hline
\end{tabular}
}
\label{table_h2_chb2}
\end{table}

\begin{table}
\centering
\caption{Same as Table~\ref{table_h2_chb1} north from the CSPN.}
\resizebox{0.35\textwidth}{!}{%
\begin{tabular}{|c|c|c|} 
 \hline
 Emission line & 1$^{\rm st}$ peak & 2$^{\rm nd}$ peak \\ [0.2ex] 
 
  & (\arcsec) & (\arcsec) \\
 \hline 
 H$_{2}$ 3.56 $\upmu$m & - & 24.88 $\pm$ 0.06 \\
 
 H$_{2}$ 2.12 $\upmu$m & - & 24.88 $\pm$ 0.03 \\
 
 H$_{2}$ 4.70 $\upmu$m & - & 24.88 $\pm$ 0.06 \\
 
 c(H~$\upbeta$) & - & 25.0 $\pm$ 0.2 \\
 \hline
\end{tabular}
}
\label{table_h2_chb3}
\end{table}


\bsp	
\label{lastpage}
\end{document}